\documentclass[useAMS,usenatbib,referee]{biom}
\usepackage{amsmath,amssymb,amsfonts,bm,amsbsy,color,bbm,mathabx}
\usepackage{multirow, verbatim}
\usepackage[figuresright]{rotating}
\usepackage{xr}
\newtheorem{pro}[theorem]{Proposition}
\newtheorem{Rem}[theorem]{Remark}

\def\bse{\begin{eqnarray*}}
	\def\ese{\end{eqnarray*}}
\def\be{\begin{eqnarray}}
	\def\ee{\end{eqnarray}}
\def\bsq{\begin{equation*}}
	\def\esq{\end{equation*}}
\def\bq{\begin{equation}}
	\def\eq{\end{equation}}
\def\bi{\begin{itemize}}
	\def\ei{\end{itemize}}

\def\sumi{\sum_{i=1}^n}

\def\sumI{\sum_{i=1}^N}

\def\wh{\widehat}
\def\wt{\widetilde}

\def\pr{\hbox{pr}}

\def\calT{{\cal T}}

\def\trans{^{\rm T}}
\def\eff{_{\rm eff}}

\def\n{\nonumber}

\def\bb{{\boldsymbol\beta}}

\def\ba{{\boldsymbol\alpha}}
\def\bphi{{\boldsymbol\phi}}

\def\0{{\bf 0}}
\def\X{{\bf X}}
\def\x{{\bf x}}

\def\c{{\bf c}}
\def\W{{\bf W}}

\def\S{{\bf S}}

\def\B{{\bf B}}
\def\b{{\bf b}}
\def\A{{\bf A}}

\def\bp{{\bf p}}

\def\ls{_{\rm ls}}

\def\calE{{\cal E}}

\def\var{\hbox{var}}
\def\cov{\hbox{cov}}

\def\boxit#1{\vbox{\hrule\hbox{\vrule\kern6pt\vbox{\kern6pt#1\kern6pt}\kern6pt\vrule}\hrule}}

\title[Guaranteed Efficiency Gain with External Dichotomized Outcomes]{Borrowing Information from an Unidentifiable Model: Guaranteed Efficiency Gain with a Dichotomized Outcome in the External Data}

\author{Lu Wang$^{1,*}$\email{lu.wang@wne.edu}, 
	Yanyuan Ma$^{2,**}$\email{yzm63@psu.edu}, and 
	Jiwei Zhao$^{3,4,***}$\email{jiwei.zhao@wisc.edu} \\
	$^{1}$Department of Mathematics, Western New England University, Springfield, Massachusetts, U.S.A \\
	$^{2}$Department of Statistics, Pennsylvania State University, University Park, Pennsylvania, U.S.A. \\
	$^{3}$Department of Statistics, University of Wisconsin-Madison, Madison, Wisconsin, U.S.A.\\
	$^{4}$Department of Biostatistics \& Medical Informatics, University of Wisconsin-Madison, Madison, Wisconsin, U.S.A.}

\usepackage{url}

\begin{document}

\date{{\it Received October} 2007. {\it Revised February} 2008.  {\it
Accepted March} 2008.}

\pagerange{\pageref{firstpage}--\pageref{lastpage}} 
\volume{64}
\pubyear{2008}
\artmonth{December}

\doi{10.1111/j.1541-0420.2005.00454.x}

\label{firstpage}

\begin{abstract}
In the era of big data, the increasing availability of diverse data
sources has driven interest in analytical approaches that integrate
information across sources to enhance statistical accuracy,
efficiency, and scientific insights. 
Many existing methods assume exchangeability among data sources and
often implicitly require that sources measure identical covariates or
outcomes, or that the error distribution is correctly
specified--assumptions that may not hold in complex real-world
scenarios. 
This paper explores the integration of data from sources with distinct
outcome scales, focusing on leveraging external data to improve
statistical efficiency. 
Specifically, we consider a scenario where the primary dataset
includes a continuous outcome, and external data provides a
dichotomized version of the same outcome. We propose two novel
estimators: the first estimator remains asymptotically consistent even
when the error distribution is potentially misspecified, while the
second estimator guarantees an efficiency gain over weighted least
squares estimation that uses the primary study data alone.
Theoretical properties of these estimators are rigorously derived, and
extensive simulation studies are conducted to highlight their
robustness and efficiency gains across various scenarios. Finally, a
real-world application using the NHANES dataset demonstrates the
practical utility of the proposed methods. 
\end{abstract}

\begin{keywords}
Data fusion, data integration, efficiency gain, efficient score, model
misspecification, unidentifiable model.
\end{keywords}

\maketitle

\section{Introduction}\label{sec:intro}

With the advent of innovative data collection methods, the growing
availability of data has sparked increasing interest in analytical
approaches that integrate data from multiple sources.
When applied effectively, integrating data and information from
diverse but relevant sources can enhance statistical accuracy, improve
efficiency, support more informed decision-making, and yield deeper
scientific insights.
In the literature, such methods and approaches are classified as data
integration \citep{lenzerini2002data} or data fusion
\citep{klein2004sensor}. 
Similarly, meta-analysis \citep{glass1976primary}, a key component of
systematic reviews, also shares a comparable spirit of combining
evidence from multiple studies. 

These methods have achieved significant success across a variety of fields.
The concept of data fusion traces back to the evolved ability of
humans and animals to incorporate information from multiple senses to
enhance survival. For instance, combining sight, touch, smell, and
taste helps determine whether a substance is edible
\citep{hall1997introduction}.
In genomics, integrating expression data, gene sequencing data, and
network data provides a heterogeneous description of genes and a
distinctive view of cellular mechanisms
\citep{lanckriet2004statistical}.
In causal inference, researchers have proposed combining data from
randomized controlled trials with observational data to evaluate the
effects of treatments or interventions on target populations different
from the study population \citep{stuart2015assessing,
  bareinboim2016causal, dahabreh2020extending, colnet2024causal}.
In machine learning, similar principles have been applied in the
contexts of semi-supervised learning, transfer learning, and
distribution shifts \citep{quinonero2008dataset}.

When combining data from different sources, assumptions are being made
about how the distributions of their corresponding populations
differ.
Many existing approaches require that these populations share a common
joint distribution or a portion of it, such as the conditional
distribution of outcome 
given covariate 
(covariate shift) or the conditional distribution of covariate
given outcome (label shift).
This concept, known as ``exchangeability'' across different
sources, enables the transfer or generalization of conclusions across
populations, thereby facilitating data integration
\citep{degtiar2023review}.

In these assumptions regarding ``exchangeability'', some more nuanced
conditions are implicitly assumed.
For example, it might be assumed that different sources
measure the exactly same set of covariate, 
or, exactly the same
outcome. 
In reality, these implicit conditions might be violated thus the
assumption regarding the common distribution cannot be justified.
\cite{li2023data} studied some data fusion techniques when different
sources of data are not perfectly aligned, and investigated the
potential efficiency gain by making use of slightly misaligned data
sources.

In this paper, we examine a situation where the outcome variables in
two data sources are on different scales.
Our motivating example involves a study where the primary outcome
variable is body mass index (BMI), defined as body mass divided by the
square of height (in kg/m$^2$).
BMI is a simple yet widely used numerical measure of a person's weight
status, enabling health professionals to discuss weight-related issues
objectively. According to the World Health Organization (WHO), adult
BMI classifications are as follows: underweight ($<$18.5 kg/m$^2$),
normal weight (18.5-24.9), overweight (25-29.9), and obese
($\geq$30). 
BMI provides a clear and practical metric for studying health outcomes.
For instance, being overweight or obese (with BMI $\geq$ 25) is
strongly associated with a variety of health problems, including but
not limited to, cardiovascular diseases, type 2 diabetes, metabolic
syndrome, mental health issues and reduced quality of life.
In our study, we incorporate an external dataset that examines the
same association between overweight status and a set of covariate
variables; 
however, this dataset only provides an indicator of
overweight (whether BMI $\geq$ 25 or not), 
rather
than the actual BMI value in the primary dataset.

The primary question of interest is how to effectively combine these
two specialized data sources to better understand the statistical
advantages of incorporating external data. 
Our initial analysis, as detailed in
Section~\ref{sec:prelim:external}, reveals that the parameter
in describing the association between the outcome and the
  covariate set
is not identifiable when relying solely on the
external data.
Consequently, the approach of 
separately estimating the parameter 
from each
data source and then combining the results is not viable. 
This raises a key question: can the external data contribute to
improving the parameter estimation,
and, if so, how?

To address this challenge, we begin by identifying all possible
estimating equations derived from the combined data. 
Specifically, as detailed in Section~\ref{sec:pro:eif}, we identify
the orthogonal complement of the nuisance tangent space,
whose elements yield regular and asymptotically
linear (RAL) estimators for the parameter of interest.
Furthermore, we derive the efficient score function 
corresponding to the semiparametric efficiency bound for
  estimating this parameter. 
To construct an estimator based on this efficient score function, we
recognize that knowledge of the error distribution in the target data
is critical. 
However, estimating this component accurately is nontrivial.
As a solution, we propose an estimator 
that uses a
potentially misspecified error distribution 
instead. 
Remarkably, the resulting locally efficient score function retains the
mean-zero property, making it a valid estimating equation. 
The corresponding estimator and its theoretical properties are
presented in Section~\ref{sec:pro:est1}. 
A minor limitation of the estimator introduced in
Section~\ref{sec:pro:est1} is that, in theory, it may not always be
more efficient than a comparable estimator based solely on the target
data, literally the weighted least squares (WLS) estimator. 
To explicitly quantify the efficiency gain from incorporating external
data, we propose a second estimator 
in Section~\ref{eff_gain}, which guarantees improved efficiency compared
to the target-data-only WLS estimator. 
Finally, in Sections~\ref{sec:sim} and \ref{sec:data}, we evaluate the
numerical performance of the proposed estimators through simulated
datasets and a real-world data application.
All the proofs are collected in the supplementary materials.

In summary, this work offers several novel contributions to the broad
field of data integration and data fusion. 
First, in our context, the parameter of interest 
is unidentifiable using external data alone; thus, a model that
integrates both target and external data is necessary. 
Second, the two proposed estimators are straightforward to implement.
Notably, they require only a potentially misspecified error
distribution 
rather than a correctly specified
and estimated one. 
This simplicity enhances their practical applicability.
The estimators also extend the same principle of the WLS estimator using target data alone. 
Finally, the second proposed estimator  
guarantees an
efficiency gain over the WLS estimator. 
This highlights the ability to effectively borrow information from an
otherwise unidentifiable model. 

\section{Problem Set-up}\label{sec:prelim}

\subsection{Parameter of interest and the target data}\label{sec:prelim:target}

In applications, investigators are usually interested in some
association between an outcome $Y$ and a set of covariates $\X$.
Here we consider $Y$ on a continuous scale which can be some
characteristic of certain disease such as clinical biomarkers and so
on.
We assume that we observe a random sample with independent and
identically distributed (i.i.d.) data $(y_i, \x_i),i=1,\ldots,n$, from
a target population $\calT$.

To study the association between $Y$ and $\X$, one may simply adopt
the linear regression model
\be\label{eq:model1}
Y = \bb\trans\X + \epsilon,
\ee
where $E(\epsilon\mid\X)=0$.
We let the first element of $\X$ to be one, so the first element of
$\bb$ is the intercept.
We assume the conditional distribution of $\epsilon$ given $\x$
follows $\epsilon\sim f(\epsilon,\x)$, where $f(\cdot)$ is an unknown
generic conditional probability
density function satisfying $\int tf(t,\x)dt=0$.
We assume that the marginal distribution of $\X$ follows $f_t(\cdot)$.
We also denote $v(\x)\equiv E(\epsilon^2\mid \x)$ and assume that $0< v(\x) < \infty$.

Based on model (\ref{eq:model1}), the simplest approach for obtaining
an estimator for $\bb$ is the ordinary least squares (OLS) estimation, that corresponds to solving the empirical version of the estimating equation $E\{(Y-\bb\trans\X)\X\}=\0$.
The OLS estimation does not need a model
specification for the heteroscedastic error distribution
$f(\epsilon,\x)$ or its conditional variance $v(\x)$.

One can cast model (\ref{eq:model1}) as a semiparametric model
regarding both $f(\epsilon,\x)$ and $f_t(\x)$ as nonparametric
nuisance; see Section 4.5 of \cite{tsiatis2006semiparametric}.
Accordingly, the efficient score for estimating $\bb$ is $\S\ls =
v(\x)^{-1}(y-\bb\trans\x)\x$.
Thus, one can obtain the corresponding estimator $\wt\bb\ls$ by solving
$\frac1n\sumi v(\x_i)^{-1}(y_i-\bb\trans\x_i)\x_i = \0$,
if $v(\x)$, or in general $f(\epsilon,\x)$, were known or could be
well estimated.
In practice, estimating $v(\x)$ or $f(\epsilon,\x)$ might not be straightforward.
Nevertheless, with a working model $f^*(\epsilon,\x)$ and thus
$v^*(\x)$, one can still obtain the estimator $\wt\bb\ls^*$ by solving
\be\label{eq:solvebblsstar}
\frac1n\sumi v^*(\x_i)^{-1}(y_i-\bb\trans\x_i)\x_i = \0.
\ee
Note that this estimator corresponds to the locally efficient score
\be\label{eq:slsstarlocal}
\S\ls^* = v^*(\x)^{-1}(y-\bb\trans\x)\x,
\ee
and we usually refer it as the WLS estimator.

The WLS estimator $\wt\bb\ls^*$ indicates
that, one is still able to obtain an asymptotically consistent
estimator even if the error distribution is misspecified.
In this paper, we are interested in estimating the parameter $\bb$
bearing in mind that the error distribution $f(\epsilon,\x)$ generally
cannot be correctly specified.
We focus on the estimator $\wt\bb\ls^*$ but also care about the
property of the estimator $\wt\bb\ls$ when $f(\epsilon,\x)$ is
correctly specified.
\begin{lemma}\label{lemma:benchmark}
	Consider model (\ref{eq:model1}) but with a possibly misspecified
	heteroscedastic error distribution $f^*(\epsilon, \x)$.
	Assume $\A_0=E(\partial\S\ls^*/\partial\bb\trans) =E\{- v^*(\X)^{-1} \X \X\trans\}$ is invertible with
	$\S\ls^*$ defined in (\ref{eq:slsstarlocal}).
	Then, the target data only estimator $\wt\bb\ls^*$ that solves (\ref{eq:solvebblsstar}) satisfies
	$n^{1/2}(\wt\bb\ls^*-\bb)\to N(\0, \A_0^{-1}\B_0{\A_0^{-1}}\trans)$,
	where $\B_0=E({\S\ls^*}^{\otimes2}) = E \{ v^*(\X)^{-2} (Y - \bb\trans \X)^2 \X \X\trans \}$.
\end{lemma}

This is the standard result about WLS
estimator hence we omit the proof.

\subsection{External data with a dichotomized outcome}\label{sec:prelim:external}

In practice, similar data sets for the same disease exist.
In this paper, other than the target population $\calT$ introduced in
Section~\ref{sec:prelim:target}, we also consider a data set with
i.i.d. random samples from an external population $\calE$.
However, it is more common that
in such a data set, the outcome variable is not $Y$ introduced above
but a dichotomous version of $Y$, the disease status.
In particular, we consider that the external population contains a random
sample $(z_i, \x_i), i=1,\ldots,m$, where $z_i=1$ if $Y_i\le c$ and
$z_i=0$ if $y_i> c$. Thus,
\be\label{eq:model2}
\pr(Z=1\mid \X)=\pr(Y\leq c \mid \X)=\pr(\epsilon\leq
c-\bb\trans\X\mid\X)=F(c-\bb\trans\X,\X),
\ee
where $F(\cdot)$ is the cumulative distribution function of $f(\cdot)$
and the cutoff value $c$ is known.
We assume in the external population the marginal distribution of $\X$ follows
$f_e(\cdot)$, which is allowed to be different from
$f_t(\cdot)$.

\begin{Rem}\label{rem:iden}
Model (\ref{eq:model2}) is semiparametric with both $F(\cdot)$ and $\bb$ unknown.
It is interesting to note that,   based on external data alone,
the parameter of interest $\bb$ in (\ref{eq:model2})  is not
identifiable, as we verify in
Section~S.1 in the supplementary materials. 
\end{Rem}

The observation in Remark~\ref{rem:iden} is critical, as it
  indicates that no estimator for $\bb$ exists by
using the external data alone.
Thus, it is infeasible to estimating $\bb$ separately from the two
data sources and then conducting meta analysis.
Nevertheless, the research goal of this paper is still to investigate
the potential benefits for estimating $\bb$, more specifically the
guaranteed efficiency gain, by incorporating this specific external data set.

\subsection{Research goal of this paper}\label{sec:prelim:goal}

For a smoother technical presentation, we pool the data from two difference sources $\calT$ and $\calE$ together, and create a binary indicator $R$ in that $R=1$ if the corresponding
subject is from the target population $\calT$ and $R=0$ if from $\calE$.
Table~\ref{tb:data} illustrates the data structure after the combination.
In this combined population $\calT \cup \calE$, we define $\pr(R=1)=\pi=n/N$ with $N=n+m$.
We also define $p(\x)\equiv E(R\mid\X)=f_t(\x)\pi/\{f_t(\x)\pi+f_e(\x)(1-\pi)\}$.

Recall that with the target data only, as explained in
Section~\ref{sec:prelim:target}, one can estimate $\bb$ via the locally efficient score $\S\ls^*$, with the
OLS as a special case (misspecifying the
conditional variance as a constant).
For later use, we rewrite the efficient score $\S\ls$ and the locally
efficient score $\S\ls^*$ as
\be
\S\ls &=& r v(\x)^{-1}(y-\bb\trans\x)\x, \mbox{ and }\label{eq:sls}\\
\S\ls^* &=& r v^*(\x)^{-1}(y-\bb\trans\x)\x, \label{eq:slsstar}
\ee
respectively, where $r$ is the realization of the random variable $R$.
Note that the estimator $\wt\bb\ls^*$ studied in
Lemma~\ref{lemma:benchmark} serves as the benchmark throughout
this paper.

Thus, the research goal of this paper is to investigate the wise use of the external data.
Compared to the estimator $\wt\bb\ls^*$ that relies on a misspecified
error distribution $f^*(\epsilon, \x)$, what can external data bring
us and what is the benefit?
More specifically, can external data enhance the estimation efficiency, compared
to $\wt\bb\ls^*$?

The answer is yes.
The information for estimating $\bb$ can be seen from the likelihood function.
Note that, while the conditional density of $Y$ given $\X$ in the
target population is $f(y-\bb\trans\x,\x)$, the conditional density of
$Z$ given $\X$ in the external population is
$\{F(c-\bb\trans\x,\x)\}^z \{1-F(c-\bb\trans\x,\x)\}^{1-z}$.
More rigorously, if we spell out the likelihood function of one
generic subject from the population $(R,RY,(1-R)Z,\X)$, which is
\be\label{eq:model}
f(y-\bb\trans\x,\x)^r [\{F(c-\bb\trans\x,\x)\}^z
\{1-F(c-\bb\trans\x,\x)\}^{1-z}]^{1-r} f_t(\x)^r
f_e(\x)^{1-r}\pi^r(1-\pi)^{1-r},
\ee
it is clear that the external data indeed bring additional information about $\bb$.

\section{Proposed Estimators}\label{sec:pro}

The key to our proposal stems from the careful investigation of model
(\ref{eq:model}) that integrates both sources of data together.
Even though the parameter of interest $\bb$ is not identifiable from
the external data only model (\ref{eq:model2}), it is from model
(\ref{eq:model}).

In Section~\ref{sec:pro:eif} we present the motivation for estimating
$\bb$ with the integrated data, where we first characterize all of the
possible mean zero estimating equations then derive the efficient
score function $\S\eff$ \citep{bickel1993efficient,
	tsiatis2006semiparametric} for estimating $\bb$ in model
(\ref{eq:model}).
In Sections~\ref{sec:pro:est1} and \ref{eff_gain} respectively, we
propose two different estimators for $\bb$.
The first more closely relies on the locally efficient score
function $\S\eff^*$, the same format as $\S\eff$ but with the same
misspecified error distribution $f^*(\epsilon,\x)$ as in the estimator
$\wt\bb\ls^*$.
Realizing that the first proposal does not always lead to
efficiency gain over the benchmark $\wt\bb\ls^*$, we propose the
second estimator in Section~\ref{eff_gain} that guarantees the
efficiency gain thus guarantees the statistical benefits by
incorporating the external data $\calE$.

\subsection{Motivation for estimating $\bb$ with the integrated
	data}\label{sec:pro:eif}

To construct estimators from the integrated data, we inspect both
  data sets separately first, then consider combining them. To this
  end, we derive the space $\Lambda^\perp$, which contains all
  possible estimating functions, and then derive the optimal member in
  $\Lambda^\perp$, which leads to the most efficient estimator.

With the target data only, 
the estimating function family can be derived to be
$\{r (y-\bb\trans\x)\c(\x), \forall \c(\x)\}$,
which contains $\S\ls$ in (\ref{eq:sls}) and $\S\ls^*$ in
(\ref{eq:slsstar}) as special elements. 
On the other hand, with the external data only, in model
(\ref{eq:model2}), if $F(\cdot)$ were known, the score function for
estimating $\bb$ is
\be\label{eq:model2score}
(1-r)\{z-F(c-\bb\trans\x,\x)\}\b_1(\x),
\ee
with
$\b_1(\x)=F(c-\bb\trans\x,\x)^{-1}\{1-F(c-\bb\trans\x,\x)\}^{-1}f(c-\bb\trans\x,\x)\x$.
Generally, for an arbitrary $\b_1(\x)$, the score function
(\ref{eq:model2score}) still maintains mean zero.

Though the score function (\ref{eq:model2score}) by itself does not
work in our setting, its format provides us some insight of how the
external data can contribute to the
integrated data model (\ref{eq:model}).
Specifically, in Proposition \ref{pro:decomp}, we derive all the
possible estimating functions.
\begin{pro}\label{pro:decomp}
	The nuisance tangent space orthogonal complement
	$\Lambda^\perp$ for estimating $\bb$ in the integrated data model
	(\ref{eq:model}) is
	\bse
	\Lambda^\perp = \left\{ r(y-\bb\trans\x)\c(\x) + \{r-p(\x)\}\{z-F(c-\bb\trans\x,\x)\}\b(\x): \forall \c(\x), \forall \b(\x) \right\}.
	\ese
\end{pro}
The structure of $\Lambda^\perp$ is important to us, since any
element in it can lead to a RAL estimator for the parameter of interest
$\bb$.
However, it does not provide us guidance on how to ``best'' choose the
functions $\c(\x)$ and $\b(\x)$.
To this end, we further derive the efficient score function $\S\eff$, which is the
projection of the score vector $\S_\bb$ onto the space
$\Lambda^\perp$.
It can lead to the most efficient estimation and will also give us
more insights on what estimators can be 
proposed, especially with a misspecified error distribution
$f^*(\epsilon,\x)$.
\begin{pro}\label{pro:eif}
	The efficient score for estimating $\bb$, $\S\eff$, is
	\bse
	\frac{r (y-\bb\trans\x)\{F(c-\bb\trans\x,\x)-1\}F(c-\bb\trans\x,\x)
		+\{r-p(\x)\} \{z-F(c-\bb\trans\x,\x)\} E(Z\epsilon\mid\x)
	}{\{E(Z\epsilon\mid\x)\}^2\{1-p(\x)\}+F(c-\bb\trans\x,\x)\{F(c-\bb\trans\x,\x)-1\}
		v(\x)}\x.
	\ese
\end{pro}
One can verify that, $\S\eff$ can be written as
\bse
\S\eff = \kappa(\x)\S\ls + \{1-\kappa(\x)\}\frac{r-p(\x)}{1-p(\x)}\frac{z-F(c-\bb\trans\x,\x)}{E(Z\epsilon\mid\x)}\x, \mbox{ where }
\ese
\bse
\kappa(\x) \equiv \frac{F(c-\bb\trans\x,\x)\{F(c-\bb\trans\x,\x)-1\}v(\x)
}{\{E(Z\epsilon\mid\x)\}^2\{1-p(\x)\}+F(c-\bb\trans\x,\x)\{F(c-\bb\trans\x,\x)-1\}
	v(\x)},
\ese
and $\S\ls$ has been defined in (\ref{eq:sls}).
Clearly, the efficient score $\S\eff$ is a weighted average between
the efficient score $\S\ls$ of model (\ref{eq:model1}) and
\be\label{eq:eff2}
\frac{r-p(\x)}{1-p(\x)}\frac{z-F(c-\bb\trans\x,\x)}{E(Z\epsilon\mid\x)}\x,
\ee
which summarizes the information in model (\ref{eq:model2}).
Note that (\ref{eq:eff2}) is not the efficient score function
(\ref{eq:model2score}), while we would expect these two
  expressions to be identical had $F(\cdot)$ been known and hence the
  model had been identifiable using the external data alone.
This difference reflects the contrast between an
unidentifiable model and an identifiable model in terms of the contributions to the
efficient score function in the integrated data model (\ref{eq:model}).

\subsection{Proposed estimator according to the efficient score
	$\S\eff$}\label{sec:pro:est1}

Our next step is to propose estimators for $\bb$ based on the efficient score $\S\eff$, especially with a misspecified error distribution $f^*(\epsilon,\x)$.
In order to present our ideas more clearly, we first consider the situation that
$f_t(\x)=f_e(\x)$; i.e., $p(\x)=\pi$, a constant.
This is a technically simpler situation.
In reality, it is also intuitive that, it is of primary interest to
first consider the situation that both sources of subjects are from
the same population.
If the statistical benefits brought by the external data are clear or
convincing in this simpler case, then we further consider a
more general situation that $f_t(\x)\neq f_e(\x)$.

When $p(\x)=\pi$, the efficient score can be simplified as
\be\label{eq:effsimple}
\S\eff = \kappa(\x)\S\ls + \{1-\kappa(\x)\}\frac{r-\pi}{1-\pi}\frac{z-F(c-\bb\trans\x,\x)}{E(Z\epsilon\mid\x)}\x.
\ee
With a possibly misspecified error distribution $f^*(\epsilon,\x)$, and subsequently $F^*(\epsilon,\x), E^*(Z\epsilon\mid\x)$,
$v^*(\x)$ and $\kappa^*(\x)$, one can obtain
\bse\label{eq:effsimple}
\S\eff^* = \kappa^*(\x)\S\ls^* + \{1-\kappa^*(\x)\}\frac{r-\pi}{1-\pi}\frac{z-F^*(c-\bb\trans\x,\x)}{E^*(Z\epsilon\mid\x)}\x, \mbox{ with }
\ese
\bse
\kappa^*(\x) = \frac{F^*(c-\bb\trans\x,\x)\{F^*(c-\bb\trans\x,\x)-1\}v^*(\x)
}{\{E^*(Z\epsilon\mid\x)\}^2\{1-\pi\}+F^*(c-\bb\trans\x,\x)\{F^*(c-\bb\trans\x,\x)-1\}
	v^*(\x)},
\ese
and $\S\ls^*$ has been defined in (\ref{eq:slsstar}).
Clearly $\S\eff^*$ preserves the mean zero property, so a corresponding RAL estimator can be proposed, with details studied later.

Further, when $p(\x)$ is not a constant, we
assume the truth of $p(\x)$ can be written as $p(\x,\ba_0)$ with $\ba_0$ the true value
of the parameter, and we assume that we are able to estimate $\ba$ such that
$N^{1/2}(\wh\ba-\ba)=N^{-1/2}\sumI \bphi(r_i,\x_i,p(\cdot),\ba)+o_p(\pi^{1/2})$.
Then, with a possibly misspecified error distribution $f^*(\epsilon,\x)$,
we have
\be\label{eq:seffs}
\wh\S\eff^*\equiv\S\eff^*(r,ry,z,\x,\bb,\wh p)
= \kappa^*(\x)\S\ls^* + \{1-\kappa^*(\x)\}\frac{r-\wh p(\x)}{1-\wh p(\x)}\frac{z-F^*(c-\bb\trans\x,\x)}{E^*(Z\epsilon\mid\x)}\x,
\ee
as the local efficient score to construct estimating equations,
where $\wh p(\x) = p(\x, \wh \ba)$, and
\bse
\kappa^*(\x) = \frac{F^*(c-\bb\trans\x,\x)\{F^*(c-\bb\trans\x,\x)-1\}v^*(\x)
}{\{E^*(Z\epsilon\mid\x)\}^2\{1-\wh p(\x)\}+F^*(c-\bb\trans\x,\x)\{F^*(c-\bb\trans\x,\x)-1\}
	v^*(\x)}.
\ese

Clearly, similar to $\S\eff$, $\wh\S\eff^*$ is also a weighted average between the locally efficient score $\S\ls^*$ of model (\ref{eq:model1}) and
\be\label{eq:eff2star}
\frac{r-\wh p(\x)}{1-\wh p(\x)}\frac{z-F^*(c-\bb\trans\x,\x)}{E^*(Z\epsilon\mid\x)}\x.
\ee
Taking into account of Remark~\ref{rem:iden},
when one models $F$
as, say, $F^*$, model (\ref{eq:model2}) becomes identifiable and the
corresponding efficient score will be
\bse
\frac{z-F^*(c-\bb\trans\x,\x)}{
	F^*(c-\bb\trans\x,\x)\{1-F^*(c-\bb\trans\x,\x)\}
}f^*(c-\bb\trans\x)\x,
\ese
which, however, is still different from (\ref{eq:eff2star}).
Similar to the discussion in the end of Section~\ref{sec:pro:eif}, this difference reflects the difference, between a posited and an unknown $F(\cdot)$ model,
of the contributions to the locally
efficient score function in the integrated data model (\ref{eq:model}).

We denote the corresponding estimators based on $\S\eff$ and
$\wh\S\eff^*$ as $\wt\bb\eff$ and $\wt\bb\eff^*$, respectively.
More specifically, $\wt\bb\eff^*$ satisfies
\be
\frac1N \sumI \S\eff^*(r_i,r_iy_i,z_i,\x_i,\wt\bb\eff^*,\wh p)=\0.
\ee
For $\wt\bb\eff^*$, 
we have the following result:
\begin{theorem}\label{th:seff}
	The estimator $\wt\bb\eff^*$ satisfies
	$n^{1/2}(\wt\bb\eff^*-\bb)\to N(\0, \A^{-1}\B{\A^{-1}}\trans)$, where 
	\be\label{eq:A}
	\A&=&E[\pi^{-1}\partial\S\eff^*\{R,RY,Z,\X,\bb,p(\X)\}/\partial\bb\trans]\n\\
	&=&E\left\{
	\frac{p(\X)\X}{\pi}E\left(\frac{\partial}{\partial\bb\trans}
	\left[\frac{ \epsilon(F^*-1)F^*
	}{\{E^*(Z\epsilon\mid\X)\}^2+F^*(F^*-1)
		E^*(\epsilon^2\mid\X)}\right]\mid\X\right)\right\}, \mbox{ and }
	\ee
	\be\label{eq:B}
	\B
	&=&E\left\{\frac{1}{\pi}\left(\S\eff^*\{R,RY,Z,\X,\bb,p(\X)\}\right.\right.\n\\
	&&\left.\left.+E\left[
	\frac{(F^*-F) E^*(Z\epsilon\mid\x)
		\x\bp_\ba'(\X,\ba)\trans}{\{E^*(Z\epsilon\mid\x)\}^2\{1-p(\x)\}+F^*(F^*-1)
		E^*(\epsilon^2\mid\x)}\right]
	\bphi(R,\X, p(\cdot),\ba)\right)^{\otimes2}\right\}.
	\ee
\end{theorem}

\begin{Rem}[Discussion when $N>>n$]
	Sometimes, the external population data may have a much
	larger sample size than the target population, we thus consider
	the case of $N>>n$ next.
	Interestingly, the much larger external data
	size does not help to accelerate the convergence rate of $\bb$, i.e., the
	convergence rate of $\wt\bb\eff^*$ remains to be $n^{-1/2}$.
In contrast, if the external data itself were
  sufficient to identify $\bb$, we would obtain convergence rate $N^{-1/2}$.
        The
	intuition behind the unimprovable convergence rate is that the external data itself does
	not suffice to identify $\bb$, as we described before.
	To capture the $N>>n$ situation, instead of $\pi, p(\x)$, we write
	$\pi_n=n/N,
	p_n(\x)=f_t(\x)\pi_n/\{f_t(\x)\pi_n+(1-\pi_n)f_e(\x)\}$, and note that
	$\pi_n\to0, p_n(\x)/\pi_n\to f_t(\x)/f_e(\x)$ when $n\to\infty$. All the results
	above hold with $\pi$ replaced by $\pi_n$.
	In this case, although we still have the form
	$N^{1/2}(\wh\ba-\ba)=N^{-1/2}\sumI\bphi(R_i,\X_i, p_n(\cdot),\ba)+o_p(\pi_n^{1/2})$,
	we note that
	$\var\{\bphi(R,\X, p_n(\cdot),\ba)^{\otimes2}\}\pi_n=\var(n^{1/2}\wh\ba)=O(1)$,
	i.e., \\
	$\var\{\bphi(R,\X, p_n(\cdot),\ba)^{\otimes2}\}=O(\pi_n^{-1})$.
\end{Rem}

\begin{Rem}[Discussion on model misspecification]
	In this Section, we advocate the modeling of $f(\epsilon,\x)$ and $p(\x)$ differently.
	The correct specification of the error distribution $f(\epsilon,\x)$
	will result in the  correct specification of the conditional
	distribution of $Y$ given $\X$ in the target data, which is a
	demanding ultimate goal.
	Given the situation that the model $f(\epsilon,\x)$ is heavily involved in implementing the efficient score function, we pursue the direction with an arbitrary working model $f^*(\epsilon,\x)$.
	On the contrary, even though it also involves the same covariate $\x$, modeling $p(\x)$ is a standard procedure for estimating a density ratio function, or, equivalently, a classification task.
	Thus, any method, either parametric or semiparametric or nonparametric, can be adopted.
	We choose a parametric approach for simplicity.
	In reality, if sufficient data are available, one may use black-box machine learning methods for estimating both $\wh f(\epsilon,\x)$ and $\wh p(\x)$ that satisfy the $o_p(n^{-1/4})$
	convergence rate in terms of the supnorm.
	In such a situation, we can engage the estimated functions in the construction of $\S\eff$ to
	achieve optimal efficiency for estimating $\bb$.
\end{Rem}

\subsection{Proposed estimator that has guaranteed efficiency gain}
\label{eff_gain}

While obviously $\wt\bb\eff$ is more efficient than $\wt\bb\ls$, we
cannot conclude $\wt\bb\eff^*$ is also always more efficient than
$\wt\bb\ls^*$. The relative performance depends on $\pi$ and the working model $f^*(\epsilon,\x)$. This is the limitation of the proposed estimator $\wt\bb\eff^*$. In this section, we propose a new estimator
\be\label{eq:pro}
\wh\bb^*=\wt\bb\ls^* + \W (\wt\bb\eff^*-\wt\bb\ls^*),
\ee
where $\W= -\{\cov(\wt\bb\ls^*, \wt\bb\eff^*-\wt\bb\ls^*)\}\{\var(\wt\bb\eff^*-\wt\bb\ls^*)\}^{-1}$.
Obviously, the estimator $\wh\bb^*$ is also asymptotically consistent, and it guarantees the safe use of the external data. Moreover, it is guaranteed to be always more efficient than $\wt\bb\ls^*$. This indicates, even with an arbitrary working model $f^*(\epsilon,\x)$, the leverage of external data is always beneficial. This optimality holds in practice as long as the unknown $\W$ is replaced by a consistent estimator $\wh\W$. Refer to Section~S.5 in the supplementary materials for a detailed discussion.

\section{Simulation Studies}\label{sec:sim}

We evaluate the numerical performance of the proposed estimators through two simulation studies. Simulation 1 assumes $f_t(\x)=f_e(\x)$ so that $p(\x) = \pi$ is a constant. Simulation 2 assumes $f_t(\x)\neq f_e(\x)$ requiring the estimation of $p(\x)$ to implement the proposed estimators. Specifically, data are generated from $Y = \beta_0 + X_1\beta_1 + X_2\beta_2 + \epsilon$, with true coefficients $(0, 1, -1)$. In simulation 1, $X_1 \sim \text{Bernoulli}(0.5)$ and $X_2 \sim \text{N}(0,1)$. In Simulation 2, the target population has $X_{1t}$ and $X_{2t}$ the same as simulation 1, while the external population has $X_{1e}\sim \text{Bernoulli}(0.3)$ and $X_{2e} \sim \text{U}(-2,2)$. For the external population, we generate a binary response $Z$ using a cutoff $c=0$ (i.e., $Z=1$ if $Y \le 0$ and $Z=0$ otherwise), resulting in approximately $60\%$ of observations having $Z=1$.
	
Both simulations examine two error distributions to assess the robustness of the proposed methods: a mixture of normals (Part 1) and a heavy-tailed $t$-distribution with 4 degrees of freedom (Part 2). Specifically, for the mixture of normals, the error density in Simulation 1 is $f(\epsilon) = 0.4\phi\{(\epsilon-8)/2\} + 0.2 \phi(\epsilon + 2)$, whereas in Simulation 2 it is $f(\epsilon) = 0.9/1.1\phi\{(\epsilon-0.2)/1.1\} + 0.1 \phi(\epsilon + 1.8)$ as the setting is more complicated. For each part, we generate $1,000$ datasets with sample sizes $n=500$ and $n=2,400$, fixing the proportions of observations from the target population at $\pi =0.5$ and  $\pi =0.25$, respectively. 

We compare the two proposed estimators $\wt\bb\eff^*$ in (\ref{eq:seffs}) and $\wh\bb^*$ in (\ref{eq:pro}) against the benchmark WLS estimator $\wt\bb\ls^*$ in (\ref{eq:slsstar}) under three scenarios: (I) correctly specified error distribution; (II) misspecification as standard normal, i.e. the working
model is
$f^*(\epsilon,\x) = (2\pi)^{-1/2} e^{-{\epsilon^2}/{2}}$
; and (III) misspecification as standard logistic, i.e.
$f^*(\epsilon,\x) = {e^{-\epsilon}}/{(1 + e^{-\epsilon} )^2}$. Note that in scenario I, the three estimators
become $\wt\bb\eff$, $\wh\bb$ and $\wt\bb\ls$, respectively. 

To address the computational challenges associated with obtaining the covariance matrix, a weighted bootstrap approach \citep{jin2001simple} has been adopted in the simulation study.
This resampling method allocates independent, identically distributed positive random weights to each data point. Subsequently, the covariance matrix is estimated by calculating the empirical covariance of the weighted estimates generated across multiple bootstrap samples.
In each of the simulation scenarios, the random weights are generated from an exponential distribution with mean of 1, and the number of bootstrap samples is fixed at 1,000. 

Additionally, we consider two benchmark methods that can be adopted
naively, but are in fact unsuitable in this context. The first benchmark, $\wh\bb_{bm1 }$,
  dichotomizes the observed response variable $y$ from the target data
  to construct a binary outcome $z$. A logistic regression model is
  then fitted on the combined target and external data using $z$ as
  the response. The second benchmark takes a meta-analytic perspective
  by analyzing the two datasets separately and then combining the
  results. Specifically, weighted least squares (WLS) is applied to
  the target data to obtain an estimator $\wh\bb_1$, while a logistic
  regression is fitted to the external data to obtain an estimator
  $\wh\bb_2$. The pooled result is then taken as the second benchmark estimator
	$\wh\bb_{bm2} = (\W_1+\W_2)^{-1}(\W_1\wh\bb_1+\W_2\wh\bb_2),$ where
	$\W_1=\var(\wh\bb_1)^{-1}$, $\W_2=\var(\wh\bb_2)^{-1}$.
	
Table~\ref{tab:mixeff} summarizes the results for Simulation 1 Part 1. Across all three scenarios, the proposed methods and the benchmark WLS estimator exhibit negligible bias, confirming consistency regardless of model specification. Regarding estimation variability in Scenario I, $\wt\bb\eff$ and $\wh\bb$ exhibit comparable performance. Indeed, since $\wt\bb\eff$ is efficient in this case, $\wt\bb\eff$ and $\wh\bb$ are asymptotically equivalent; by leveraging information from both the target and external populations, they achieve greater efficiency than $\wt\bb\ls$, which relies solely on target data. In Scenarios II and III, $\wt\bb\eff^*$ is no longer guaranteed to be efficient. However, in consonance with our expectation, $\wh\bb^*$ shows an advantage in estimation variability, yielding smaller SSDs than both $\wt\bb\ls^*$ and $\wt\bb\eff^*$. Finally, all the methods have sample
standard deviation (SSD) close to their corresponding average
estimated standard error (ESE), and
the 95\% coverage
rates of all the proposed methods are around the nominal level.
This indicates that the estimated
standard errors in the proposed methods are sufficiently
precise. In general, this simulation study suggests that all three methods demonstrate consistency in estimating the regression coefficients regardless the working model is misspecified or not. It is worth noting that generally $\wt\bb\eff^*$ outperforms $\wt\bb\ls^*$, and when the working model is misspecified, the weighting strategy can further improve the performance of $\wt\bb\eff^*$. In a separate note, two benchmark methods ($\widehat{\boldsymbol{\beta}}_{bm1}$ and $\widehat{\boldsymbol{\beta}}_{bm2}$) yield substantial biases and poor coverage rates.

The results for Simulation 1 Part 2 are summarized in Table~\ref{tab:t}. Consistent with the findings in Part 1, our proposed estimators remain asymptotically unbiased and perform well across all scenarios, demonstrating strong robustness to heavy-tailed errors even when the working model is misspecified.

Different from Simulation 1, in Simulation 2,
$p(\x)$ is estimated parametrically in all three
scenarios. Specifically, we use $\wh p(\x) = \wh f_t(\x)\pi/\{\wh f_t(\x)\pi+ \wh
f_e(\x)(1-\pi)\}$, where $\wh{f_t}(\x) = s^{-1}_{x_{2t} }\phi\{(x_2 -
\overline{x}_{2t})/s_{x_{2t}}\}(\overline{x}_{1t})^{x_1}(1-\overline{x}_{1t})^{(1-x_1)}$
and $\wh{f_e}(\x) = [1/\{\max (x_{2e}) - \min
(x_{2e})\}](\overline{x}_{1e})^{x_1}(1-\overline{x}_{1e})^{(1-x_1)}$.
Here, $\overline x_{1e}, \overline x_{1t}$, $\overline x_{2t}$, and
$s_{x_{2t}}$ are respectively the sample means of $x_1$ in the
external population, in the target population, and the sample mean and
sample standard deviation of $x_2$ in the target population.
The results are summarized in Table \ref{tab:mixeff_2} and Table \ref{tab:t2} for Part 1 and 2, respectively.
Briefly, the conclusions from Simulation 2 are the same as Simulation 1.
It is clear that when covariates in target population and external population have different distributions, the proposed methods are all asymptotically unbiased. Meanwhile, most of the time $\wt\bb\eff^*$ is more precise than $\wt\bb\ls^*$, and $\wh\bb^*$ consistently maintains higher precision than $\wt\bb\ls^*$, aligning with our theory.

\section{Real Data Application}\label{sec:data}


The National Health and Nutrition Examination Survey (NHANES) is a program designed by the National Center for Health Statistics (NCHS), a division of the Centers for Disease Control and Prevention (CDC).
The NHANES collects extensive health and nutritional information from a diverse U.S. population and aims to assess the health and nutritional status of adults and children in the United States.
NHANES began conducting health and nutrition surveys in the 1960s and became a continuous program in 1999.
Annually, around 5,000 individuals of various ages are interviewed in their homes and undergo health examinations.
The comprehensive data combines survey interviews with physical examinations and laboratory tests.
It offers valuable insights contributing significantly to public health in the United States.

In this Section, we apply the proposed methods in a subset of the NHANES database previously analyzed in \cite{dinh2019data}.
This dataset contains 2,278 observations and 9 variables: age group (senior/non-senior), age, gender, engagement in moderate or vigorous-intensity sports, fitness, or recreational activities during a typical week (PAQ605), body mass index (BMXBMI), blood glucose after fasting (LBXGLU), diabetes status (DIQ010), oral glucose tolerance test (LBXGLT), and blood insulin levels (LBXIN).

The primary objective of this analysis is to investigate the impact of
various predictors (Age, Gender, PAQ605, LBXGLU, LBXGLT, LBXIN) on
Body Mass Index (BMXBMI). One observation with erroneous values for
PAQ605 and 13 observations with extreme covariate values (i.e. at
least one of the covariates is more than 4 standard deviations away
from the mean), have been excluded from the analysis. This results in a dataset containing 2,264 observations. The target population consists of 574 randomly selected individuals, whereas the remaining 1,690 individuals form the external population. As previously described, the dataset structure ensures that observations within the target population contain precise BMI values. In contrast, for the external group, BMI is categorized into overweight status: a BMI below 25 is classified as 1 (not overweight), and a BMI of 25 or higher is classified as 0 (overweight).

Prior to model fitting, all numerical predictors have been
logarithmized, centered, and scaled. The error term in the working
model is assumed to follow a standard logistic distribution and then a
standard normal distribution for comparison purposes. The weighted
bootstrap approach, as mentioned in the simulation study, has been
adopted to facilitate the statistical inference, with the number of
bootstrap sample set at 1,000.
Additionally, we implement the two benchmark methods described in the simulation studies, $\widehat{\boldsymbol{\beta}}_{bm1}$ and $\widehat{\boldsymbol{\beta}}_{bm2}$, as a comparison.
The results from assuming logistic error, the normal error and the two benchmark methods are summarized in Table~\ref{tab:bmi}.

The results in Table~\ref{tab:bmi} implying that the proposed methods have reliable performance across different working models. It is also worth noting that $\wh{\bb}^*$ consistently be more efficient compared to the other methods, as evidenced by its lower estimated standard errors. The analysis indicates a positive association between Age and BMI across all methods, suggesting an increase in BMI with age. Gender is found to be a non-significant predictor showing a negligible effect. Physical activity (PAQ605) demonstrates a consistent negative relationship with BMI, indicating that higher activity levels are associated with lower BMI values. This relationship is not significant in the WLS estimate using data solely from the target population, but is identified as significant after incorporating external data. Biochemical markers glucose (LBXGLU) and oral glucose tolerance test (LBXGLT) exhibit non-significant associations with BMI. Notably, insulin levels (LBXIN) are positively associated with BMI across all models, highlighting a potential link between insulin resistance and increased body mass. Consistent with our simulation findings, the estimates from the two benchmark methods deviate substantially from the proposed estimators and contradict established medical knowledge. This further confirms the practical value of our proposed methods.


\backmatter


\section*{Acknowledgements}

The authors would like to thank the Co-Editor, an Associate Editor, and the two reviewers for their insightful comments which have helped improve the manuscript substantially. The research is supported in part by NSF (DMS 1953526, 2122074, 2310942), NIH (R01DC021431, R01LM014401, R01NS131225) and the American Family Funding Initiative of UW-Madison. \vspace*{-8pt}


\section*{Supplementary Materials}

Supplementary materials referenced in Sections~\ref{sec:intro}, \ref{sec:prelim:external}, and \ref{eff_gain}, and code for implementing the proposed estimators, are available with this paper at the Biometrics website on Oxford Academic. \vspace*{-8pt}

\section*{Data Availability}

Data analyzed in Section~\ref{sec:data} is available from the website \url{https://archive.ics.uci.edu/dataset/887/national+health+and+nutrition+health+survey+2013-2014+(nhanes)+age+prediction+subset}. \vspace*{-8pt}


%
\bibliographystyle{biom} 
\bibliography{refMIXEDtransfer.bib}









\begin{table}[tbp]
	\centering
	\caption{Data structure after combining two different data sources.}
	\begin{tabular}{c|c|cccc}
		\hline
		&   & $R$ & $Y$ & $Z$ & $\X$\\
		\hline
		\multirow{4}{*}{Target Data $\calT$} & 1 & 1 & \checkmark & & \checkmark\\
		& 2 & 1 & \checkmark & & \checkmark\\
		& \vdots & 1 & \checkmark & & \checkmark\\
		& $n$ & 1 & \checkmark & & \checkmark\\
		\hline
		\multirow{3}{*}{External Data $\calE$} & $n+1$ & 0 & & \checkmark & \checkmark\\
		& \vdots & 0 &  & \checkmark & \checkmark \\
		& $n+m\equiv N$ & 0 &  & \checkmark & \checkmark \\
		\hline
	\end{tabular} \label{tb:data}
\end{table}

\begin{sidewaystable}[tbp]
	\centering
	\caption{Simulation 1 Part 1. Error distribution is mixed normal. Scenario I: error distribution is correctly specified; Scenario II: misspecified as standard normal; Scenario III: misspecified as standard logistic. Results include the bias (Bias),the sample standard deviation (SSD), the average estimated standard error (ESE), and the coverage rate of 95\% confidence intervals (CR95) for the 1000 estimates.}
	\begin{tabular}{lll*{11}{r}}
		\hline
		\multicolumn{3}{c}{} & \multicolumn{3}{c}{Scenario I} & \multicolumn{3}{c}{Scenario II} & \multicolumn{3}{c}{Scenario III} & \multicolumn{2}{c}{} \\
		\cline{4-6} \cline{7-9} \cline{10-12} \vspace{-0.8em} \\
		$n$ &  &  & $\wt\bb\ls$ & $\wt\bb\eff$ & $\wh\bb$ & $\wt\bb\ls^*$ & $\wt\bb\eff^*$ & $\wh\bb^*$ & $\wt\bb\ls^*$ & $\wt\bb\eff^*$ & $\wh\bb^*$ & $\wh\bb_{bm1}$ & $\wh\bb_{bm2}$ \vspace{0.2em} \\
		\hline \vspace{-0.8em} \\
		500 & $\beta_0$ & Bias & -0.004 & 0.002 & 0.014 & 0.007 & 0.006 & 0.012 & -0.004 & -0.006 & -0.004 & -0.669 & -0.532 \\
		&  & SSD  & 0.413 & 0.365 & 0.362 & 0.397 & 0.366 & 0.349 & 0.413 & 0.366 & 0.343 & 0.137 & 0.169 \\
		&  & ESE  & 0.390 & 0.351 & 0.347 & 0.390 & 0.363 & 0.339 & 0.390 & 0.348 & 0.328 & 0.138 & 0.031 \\
		&  & CR95 & 0.932 & 0.941 & 0.941 & 0.941 & 0.939 & 0.932 & 0.932 & 0.935 & 0.931 & 0.002 & 0.003 \\
		& $\beta_1$ & Bias & 0.019 & 0.020 & 0.023 & -0.010 & -0.004 & 0.018 & 0.019 & 0.023 & 0.030 & 1.504 & 1.213 \\
		&  & SSD  & 0.575 & 0.524 & 0.528 & 0.579 & 0.537 & 0.519 & 0.575 & 0.515 & 0.495 & 0.192 & 0.239 \\
		&  & ESE  & 0.552 & 0.502 & 0.496 & 0.553 & 0.514 & 0.487 & 0.552 & 0.494 & 0.474 & 0.190 & 0.059 \\
		&  & CR95 & 0.938 & 0.933 & 0.927 & 0.940 & 0.936 & 0.937 & 0.938 & 0.936 & 0.933 & 0.000 & 0.000 \\
		& $\beta_2$ & Bias & 0.026 & 0.026 & 0.013 & 0.007 & 0.003 & -0.031 & 0.026 & 0.017 & 0.005 & -1.505 & -1.192 \\
		&  & SSD  & 0.276 & 0.257 & 0.261 & 0.284 & 0.260 & 0.259 & 0.276 & 0.248 & 0.242 & 0.104 & 0.117 \\
		&  & ESE  & 0.276 & 0.253 & 0.248 & 0.274 & 0.255 & 0.243 & 0.276 & 0.248 & 0.238 & 0.100 & 0.016 \\
		&  & CR95 & 0.949 & 0.943 & 0.933 & 0.932 & 0.934 & 0.914 & 0.949 & 0.946 & 0.945 & 0.000 & 0.000 \\
		\\
		2400 & $\beta_0$ & Bias & -0.005 & -0.030 & -0.039 & -0.006 & -0.006 & -0.003 & 0.006 & 0.005 & 0.005 & -0.664 & -0.614 \\
		&  & SSD  & 0.256 & 0.208 & 0.206 & 0.262 & 0.236 & 0.222 & 0.255 & 0.215 & 0.195 & 0.063 & 0.068 \\
		&  & ESE  & 0.254 & 0.218 & 0.209 & 0.254 & 0.229 & 0.215 & 0.253 & 0.212 & 0.192 & 0.063 & 0.005 \\
		&  & CR95 & 0.950 & 0.968 & 0.959 & 0.949 & 0.952 & 0.945 & 0.944 & 0.942 & 0.935 & 0.000 & 0.000 \\
		& $\beta_1$ & Bias & 0.017 & 0.038 & 0.048 & 0.019 & 0.017 & 0.023 & -0.001 & 0.001 & 0.005 & 1.501 & 1.394 \\
		&  & SSD  & 0.365 & 0.320 & 0.327 & 0.369 & 0.332 & 0.321 & 0.365 & 0.312 & 0.294 & 0.087 & 0.093 \\
		&  & ESE  & 0.358 & 0.340 & 0.328 & 0.358 & 0.325 & 0.310 & 0.358 & 0.302 & 0.283 & 0.086 & 0.009 \\
		&  & CR95 & 0.939 & 0.959 & 0.945 & 0.938 & 0.942 & 0.933 & 0.945 & 0.935 & 0.934 & 0.000 & 0.000 \\
		& $\beta_2$ & Bias & -0.003 & 0.005 & 0.011 & 0.003 & 0.000 & -0.023 & -0.009 & -0.008 & -0.013 & -1.500 & -1.382 \\
		&  & SSD  & 0.177 & 0.157 & 0.156 & 0.176 & 0.157 & 0.157 & 0.184 & 0.157 & 0.150 & 0.045 & 0.049 \\
		&  & ESE  & 0.178 & 0.173 & 0.157 & 0.179 & 0.161 & 0.155 & 0.178 & 0.151 & 0.142 & 0.045 & 0.003 \\
		&  & CR95 & 0.952 & 0.950 & 0.939 & 0.951 & 0.959 & 0.944 & 0.940 & 0.939 & 0.937 & 0.000 & 0.000 \\
		\hline
	\end{tabular}
	\label{tab:mixeff}
\end{sidewaystable}

\begin{sidewaystable}[tbp]
	\centering
	\caption{Simulation 1 Part 2. Error distribution is t-distribution with 4 degree of freedom. Scenario I: error distribution is correctly specified; Scenario II: misspecified as standard normal; Scenario III: misspecified as standard logistic. Results include the bias (Bias),the sample standard deviation (SSD), the average estimated standard error (ESE), and the coverage rate of 95\% confidence intervals (CR95) for the 1000 estimates.}
	\begin{tabular}{lll*{11}{r}}
		\hline
		\multicolumn{3}{c}{} & \multicolumn{3}{c}{Scenario I} & \multicolumn{3}{c}{Scenario II} & \multicolumn{3}{c}{Scenario III} & \multicolumn{2}{c}{} \\
		\cline{4-6} \cline{7-9} \cline{10-12} \vspace{-0.8em} \\
		$n$ &  &  & $\wt\bb\ls$ & $\wt\bb\eff$ & $\wh\bb$ & $\wt\bb\ls^*$ & $\wt\bb\eff^*$ & $\wh\bb^*$ & $\wt\bb\ls^*$ & $\wt\bb\eff^*$ & $\wh\bb^*$ & $\wh\bb_{bm1}$ & $\wh\bb_{bm2}$ \vspace{0.2em} \\
		\hline \vspace{-0.8em} \\
		500 & $\beta_0$ & Bias & -0.009 & -0.006 & -0.006 & -0.009 & -0.007 & -0.006 & -0.009 & -0.006 & -0.007 & 0.008 & 0.003 \\
		&  & SSD  & 0.126 & 0.112 & 0.112 & 0.126 & 0.113 & 0.112 & 0.126 & 0.114 & 0.113 & 0.151 & 0.107 \\
		&  & ESE  & 0.123 & 0.109 & 0.108 & 0.123 & 0.109 & 0.108 & 0.123 & 0.111 & 0.109 & 0.150 & 0.012 \\
		&  & CR95 & 0.947 & 0.943 & 0.939 & 0.947 & 0.941 & 0.936 & 0.947 & 0.950 & 0.939 & 0.954 & 0.163 \\
		& $\beta_1$ & Bias & 0.009 & 0.005 & 0.001 & 0.009 & 0.006 & 0.001 & 0.009 & 0.004 & 0.002 & 2.449 & 0.537 \\
		&  & SSD  & 0.174 & 0.156 & 0.157 & 0.174 & 0.157 & 0.157 & 0.174 & 0.159 & 0.158 & 0.243 & 0.169 \\
		&  & ESE  & 0.176 & 0.156 & 0.155 & 0.176 & 0.157 & 0.155 & 0.176 & 0.159 & 0.156 & 0.236 & 0.025 \\
		&  & CR95 & 0.953 & 0.957 & 0.952 & 0.953 & 0.955 & 0.950 & 0.953 & 0.965 & 0.957 & 0.000 & 0.000 \\
		& $\beta_2$ & Bias & -0.005 & -0.004 & 0.000 & -0.005 & -0.005 & 0.000 & -0.005 & -0.003 & 0.001 & -2.460 & -0.359 \\
		&  & SSD  & 0.088 & 0.081 & 0.082 & 0.088 & 0.082 & 0.083 & 0.088 & 0.082 & 0.082 & 0.155 & 0.110 \\
		&  & ESE  & 0.088 & 0.080 & 0.079 & 0.088 & 0.080 & 0.079 & 0.088 & 0.081 & 0.079 & 0.151 & 0.007 \\
		&  & CR95 & 0.954 & 0.957 & 0.952 & 0.954 & 0.954 & 0.946 & 0.954 & 0.960 & 0.952 & 0.000 & 0.000 \\
		\\
		2400 & $\beta_0$ & Bias & 0.000 & 0.001 & 0.001 & 0.000 & 0.000 & 0.000 & 0.000 & 0.001 & 0.001 & -0.002 & -0.001 \\
		&  & SSD  & 0.080 & 0.065 & 0.065 & 0.080 & 0.067 & 0.067 & 0.080 & 0.068 & 0.066 & 0.068 & 0.057 \\
		&  & ESE  & 0.081 & 0.066 & 0.066 & 0.081 & 0.067 & 0.066 & 0.081 & 0.068 & 0.066 & 0.068 & 0.003 \\
		&  & CR95 & 0.939 & 0.951 & 0.948 & 0.939 & 0.940 & 0.943 & 0.939 & 0.947 & 0.945 & 0.955 & 0.094 \\
		& $\beta_1$ & Bias & -0.003 & -0.005 & -0.005 & -0.003 & -0.004 & -0.005 & -0.003 & -0.006 & -0.005 & 2.440 & 1.136 \\
		&  & SSD  & 0.116 & 0.097 & 0.097 & 0.116 & 0.098 & 0.099 & 0.116 & 0.100 & 0.097 & 0.113 & 0.108 \\
		&  & ESE  & 0.115 & 0.095 & 0.095 & 0.115 & 0.096 & 0.095 & 0.115 & 0.098 & 0.095 & 0.107 & 0.007 \\
		&  & CR95 & 0.945 & 0.944 & 0.943 & 0.945 & 0.947 & 0.941 & 0.945 & 0.942 & 0.945 & 0.000 & 0.000 \\
		& $\beta_2$ & Bias & 0.001 & 0.001 & 0.002 & 0.001 & 0.001 & 0.001 & 0.001 & 0.002 & 0.003 & -2.441 & -0.849 \\
		&  & SSD  & 0.059 & 0.049 & 0.049 & 0.059 & 0.050 & 0.050 & 0.059 & 0.050 & 0.049 & 0.069 & 0.089 \\
		&  & ESE  & 0.057 & 0.049 & 0.049 & 0.057 & 0.050 & 0.049 & 0.057 & 0.050 & 0.049 & 0.068 & 0.002 \\
		&  & CR95 & 0.940 & 0.946 & 0.943 & 0.940 & 0.946 & 0.949 & 0.940 & 0.942 & 0.945 & 0.000 & 0.000 \\
		\hline
	\end{tabular}
	\label{tab:t}
\end{sidewaystable}

\begin{sidewaystable}[tbp]
	\centering
	\caption{Simulation 2 Part 1. Error distribution is
		mixture of normal. Scenario I: error distribution is correctly specified; Scenario II: misspecified as standard normal; Scenario III: misspecified as standard logistic. Results include the bias (Bias),the sample
		standard deviation (SSD), the average estimated standard
		error (ESE), and the coverage rate of 95\% confidence
		intervals (CR95) for the 1000 estimates.}
	\begin{tabular}{lll*{11}{r}}
		\hline
		\multicolumn{3}{c}{} & \multicolumn{3}{c}{Scenario I} & \multicolumn{3}{c}{Scenario II} & \multicolumn{3}{c}{Scenario III} & \multicolumn{2}{c}{} \\
		\cline{4-6} \cline{7-9} \cline{10-12} \vspace{-0.8em} \\
		$n$ &  &  & $\wt\bb\ls$ & $\wt\bb\eff$ & $\wh\bb$ & $\wt\bb\ls^*$ & $\wt\bb\eff^*$ & $\wh\bb^*$ & $\wt\bb\ls^*$ & $\wt\bb\eff^*$ & $\wh\bb^*$ & $\wh\bb_{bm1}$ & $\wh\bb_{bm2}$ \vspace{0.2em} \\
		\hline \vspace{-0.8em} \\
		500 & $\beta_0$ & Bias & -0.003 & -0.004 & -0.003 & -0.005 & -0.005 & -0.005 & -0.003 & -0.001 & -0.002 & -0.678 & -0.580 \\
		&  & SSD  & 0.114 & 0.099 & 0.099 & 0.110 & 0.092 & 0.092 & 0.114 & 0.097 & 0.095 & 0.129 & 0.150 \\
		&  & ESE  & 0.110 & 0.096 & 0.092 & 0.111 & 0.093 & 0.093 & 0.110 & 0.096 & 0.092 & 0.126 & 0.023 \\
		&  & CR95 & 0.941 & 0.936 & 0.928 & 0.953 & 0.950 & 0.948 & 0.934 & 0.944 & 0.937 & 0.000 & 0.000 \\
		& $\beta_1$ & Bias & -0.003 & -0.004 & -0.006 & 0.008 & 0.007 & 0.002 & 0.000 & 0.001 & -0.003 & 1.515 & 1.182 \\
		&  & SSD  & 0.156 & 0.139 & 0.138 & 0.158 & 0.138 & 0.139 & 0.160 & 0.141 & 0.140 & 0.192 & 0.255 \\
		&  & ESE  & 0.156 & 0.141 & 0.137 & 0.157 & 0.138 & 0.137 & 0.156 & 0.140 & 0.136 & 0.193 & 0.067 \\
		&  & CR95 & 0.950 & 0.950 & 0.945 & 0.948 & 0.951 & 0.952 & 0.942 & 0.947 & 0.940 & 0.000 & 0.001 \\
		& $\beta_2$ & Bias & 0.000 & -0.002 & 0.005 & -0.003 & -0.003 & -0.000 & -0.002 & -0.001 & 0.001 & -1.503 & -1.120 \\
		&  & SSD  & 0.081 & 0.080 & 0.078 & 0.077 & 0.071 & 0.071 & 0.079 & 0.074 & 0.074 & 0.114 & 0.137 \\
		&  & ESE  & 0.078 & 0.077 & 0.072 & 0.078 & 0.072 & 0.072 & 0.077 & 0.073 & 0.071 & 0.106 & 0.019 \\
		&  & CR95 & 0.930 & 0.933 & 0.921 & 0.946 & 0.944 & 0.944 & 0.940 & 0.936 & 0.931 & 0.000 & 0.000 \\
		\\
		2400 & $\beta_0$ & Bias & -0.002 & -0.003 & -0.003 & -0.003 & -0.003 & -0.003 & -0.003 & -0.002 & -0.003 & -0.671 & -0.633 \\
		&  & SSD  &  0.069 & 0.055 & 0.055 & 0.073 & 0.056 & 0.056 & 0.073 & 0.058 & 0.056 & 0.056 & 0.062 \\
		&  & ESE  & 0.072 & 0.056 & 0.055 & 0.072 & 0.055 & 0.055 & 0.072 & 0.058 & 0.055 & 0.055 & 0.004 \\
		&  & CR95 & 0.957 & 0.957 & 0.957 & 0.940 & 0.946 & 0.945 & 0.940 & 0.949 & 0.938 & 0.000 & 0.000 \\
		& $\beta_1$ & Bias & 0.005 & 0.002 & 0.002 & 0.004 & 0.003 & 0.001 & 0.004 & 0.001 & 0.001 & 1.503 & 1.378 \\
		&  & SSD  & 0.097 & 0.079 & 0.079 & 0.104 & 0.083 & 0.083 & 0.104 & 0.084 & 0.083 & 0.089 & 0.106 \\
		&  & ESE  & 0.101 & 0.083 & 0.082 & 0.101 & 0.082 & 0.081 & 0.101 & 0.085 & 0.081 & 0.090 & 0.011 \\
		&  & CR95 & 0.956 & 0.962 & 0.959 & 0.934 & 0.945 & 0.942 & 0.934 & 0.951 & 0.943 & 0.000 & 0.000 \\
		& $\beta_2$ & Bias & -0.003 & -0.003 & -0.002 & -0.000 & -0.001 & 0.000 & -0.000 & 0.000 & 0.001 & -1.484 & -1.335 \\
		&  & SSD  & 0.050 & 0.046 & 0.045 & 0.054 & 0.048 & 0.048 & 0.054 & 0.048 & 0.047 & 0.052 & 0.057 \\
		&  & ESE  & 0.051 & 0.047 & 0.045 & 0.050 & 0.045 & 0.044 & 0.050 & 0.046 & 0.045 & 0.049 & 0.003 \\
		&  & CR95 &  0.947 & 0.957 & 0.950 & 0.927 & 0.934 & 0.933 & 0.927 & 0.944 & 0.942 & 0.000 & 0.000 \\
		\hline
	\end{tabular}
	\label{tab:mixeff_2}
\end{sidewaystable}

\begin{sidewaystable}[tbp]
	\centering
	\caption{Simulation 2 Part 2. Error distribution is t-distribution with 4 degree of freedom. Scenario I: error distribution is correctly specified; Scenario II: misspecified as standard normal; Scenario III: misspecified as standard logistic. Results include the bias (Bias),the sample standard deviation (SSD), the average estimated standard error (ESE), and the coverage rate of 95\% confidence intervals (CR95) for the 1000 estimates}
	\begin{tabular}{lll*{11}{r}}
		\hline
		\multicolumn{3}{c}{} & \multicolumn{3}{c}{Scenario I} & \multicolumn{3}{c}{Scenario II} & \multicolumn{3}{c}{Scenario III} & \multicolumn{2}{c}{} \\
		\cline{4-6} \cline{7-9} \cline{10-12} \vspace{-0.8em} \\
		$n$ &  &  & $\wt\bb\ls$ & $\wt\bb\eff$ & $\wh\bb$ & $\wt\bb\ls^*$ & $\wt\bb\eff^*$ & $\wh\bb^*$ & $\wt\bb\ls^*$ & $\wt\bb\eff^*$ & $\wh\bb^*$ & $\wh\bb_{bm1}$ & $\wh\bb_{bm2}$ \vspace{0.2em} \\
		\hline \vspace{-0.8em} \\
		500 & $\beta_0$ & Bias & -0.002 & -0.001 & -0.001 & -0.002 & -0.001 & -0.001 & -0.002 & -0.002 & -0.002 & -0.002 & 0.001 \\
		&  & SSD  & 0.127 & 0.110 & 0.110 & 0.127 & 0.111 & 0.110 & 0.127 & 0.113 & 0.111 & 0.138 & 0.102 \\
		&  & ESE  & 0.124 & 0.108 & 0.107 & 0.124 & 0.109 & 0.107 & 0.124 & 0.111 & 0.108 & 0.136 & 0.011 \\
		&  & CR95 & 0.941 & 0.949 & 0.948 & 0.941 & 0.948 & 0.945 & 0.941 & 0.951 & 0.948 & 0.950 & 0.160 \\
		& $\beta_1$ & Bias & 0.003 & 0.001 & -0.004 & 0.003 & 0.002 & -0.004 & 0.003 & 0.002 & -0.002 & 2.465 & 0.469 \\
		&  & SSD  & 0.178 & 0.161 & 0.162 & 0.178 & 0.162 & 0.163 & 0.178 & 0.164 & 0.163 & 0.253 & 0.165 \\
		&  & ESE  & 0.177 & 0.158 & 0.157 & 0.177 & 0.159 & 0.157 & 0.177 & 0.161 & 0.158 & 0.242 & 0.026 \\
		&  & CR95 & 0.955 & 0.945 & 0.946 & 0.955 & 0.953 & 0.948 & 0.955 & 0.949 & 0.942 & 0.000 & 0.002 \\
		& $\beta_2$ & Bias & -0.004 & -0.005 & -0.003 & -0.004 & -0.005 & -0.002 & -0.004 & -0.004 & -0.002 & -2.467 & -0.377 \\
		&  & SSD  & 0.093 & 0.089 & 0.088 & 0.093 & 0.089 & 0.089 & 0.093 & 0.090 & 0.089 & 0.152 & 0.107 \\
		&  & ESE  & 0.087 & 0.082 & 0.081 & 0.087 & 0.082 & 0.081 & 0.087 & 0.083 & 0.081 & 0.149 & 0.007 \\
		&  & CR95 & 0.942 & 0.937 & 0.930 & 0.942 & 0.937 & 0.930 & 0.942 & 0.926 & 0.932 & 0.000 & 0.000 \\
		\\
		2400 & $\beta_0$ & Bias & 0.002 & 0.002 & 0.002 & -0.001 & -0.003 & -0.003 & 0.002 & 0.002 & 0.002 & 0.000 & 0.000 \\
		&  & SSD  & 0.084 & 0.069 & 0.068 & 0.081 & 0.067 & 0.066 & 0.084 & 0.071 & 0.069 & 0.058 & 0.050 \\
		&  & ESE  & 0.082 & 0.066 & 0.066 & 0.081 & 0.067 & 0.066 & 0.082 & 0.069 & 0.066 & 0.059 & 0.003 \\
		&  & CR95 & 0.944 & 0.950 & 0.948 & 0.959 & 0.960 & 0.954 & 0.944 & 0.943 & 0.942 & 0.953 & 0.080 \\
		& $\beta_1$ & Bias & 0.000 & -0.001 & -0.002 & 0.002 & 0.005 & 0.004 & 0.000 & -0.002 & -0.002 & 2.443 & 1.034 \\
		&  & SSD  & 0.113 & 0.096 & 0.096 & 0.111 & 0.093 & 0.092 & 0.113 & 0.099 & 0.097 & 0.114 & 0.117 \\
		&  & ESE  & 0.115 & 0.097 & 0.096 & 0.115 & 0.098 & 0.097 & 0.115 & 0.100 & 0.097 & 0.112 & 0.008 \\
		&  & CR95 & 0.953 & 0.960 & 0.957 & 0.958 & 0.963 & 0.960 & 0.953 & 0.955 & 0.954 & 0.000 & 0.000 \\
		& $\beta_2$ & Bias & -0.001 & 0.000 & -0.001 & 0.000 & 0.002 & 0.002 & -0.001 & 0.000 & 0.000 & -2.447 & -0.886 \\
		&  & SSD  & 0.062 & 0.055 & 0.055 & 0.061 & 0.055 & 0.055 & 0.062 & 0.056 & 0.055 & 0.065 & 0.100 \\
		&  & ESE  & 0.058 & 0.052 & 0.051 & 0.057 & 0.052 & 0.051 & 0.058 & 0.053 & 0.052 & 0.067 & 0.002 \\
		&  & CR95 & 0.927 & 0.930 & 0.929 & 0.937 & 0.935 & 0.935 & 0.927 & 0.935 & 0.930 & 0.000 & 0.000 \\
		\hline
	\end{tabular}
	\label{tab:t2}
\end{sidewaystable}

\begin{table}[tbp]
	\centering
	\caption{Combined summary of analytical results for the NHANES dataset. The table compares the proposed estimators ($\tilde{\boldsymbol{\beta}}_{ls}^*$, $\tilde{\boldsymbol{\beta}}_{eff}^*$ and $\widehat{\boldsymbol{\beta}}^*$) under both Logistic and Normal working error distributions, and the two benchmark methods ($\widehat{\boldsymbol{\beta}}_{bm1}$ and $\widehat{\boldsymbol{\beta}}_{bm2}$). Results include estimates (Est), estimated standard errors (ESE), and 95\% confidence intervals (CI95) for six regression coefficients.}
	\label{tab:nhanes_combined}
	\resizebox{\textwidth}{!}{%
		\begin{tabular}{llrrrrrrr}
			\hline
			& & Intercept & Age & Gender & PAQ605 & LBXGLU & LBXGLT & LBXIN \\ 
			\hline
			
			\multicolumn{9}{l}{\textit{Proposed Methods (Logistic Error Assumption)}} \\
			$\tilde{\boldsymbol{\beta}}_{ls}^*$ & Est & 28.083 & 1.949 & 0.897 & -1.107 & -0.349 & -0.118 & 4.012 \\
			& ESE & 0.496 & 0.242 & 0.469 & 0.567 & 0.343 & 0.273 & 0.269 \\
			& LowerCI95 & 27.111 & 1.474 & -0.023 & -2.219 & -1.020 & -0.653 & 3.485 \\
			& UpperCI95 & 29.056 & 2.424 & 1.817 & 0.004 & 0.323 & 0.418 & 4.539 \\
			$\tilde{\boldsymbol{\beta}}_{eff}^*$ & Est & 26.414 & 1.341 & 0.558 & -1.158 & -0.433 & -0.086 & 2.920 \\ 
			& ESE & 0.405 & 0.190 & 0.373 & 0.474 & 0.285 & 0.229 & 0.234 \\ 
			& Lower CI95 & 25.619 & 0.969 & -0.172 & -2.087 & -0.992 & -0.535 & 2.461 \\ 
			& Upper CI95 & 27.208 & 1.713 & 1.288 & -0.230 & 0.126 & 0.363 & 3.378 \\ 
			$\widehat{\boldsymbol{\beta}}^*$ & Est & 26.722 & 1.520 & 0.059 & -1.704 & -0.513 & -0.123 & 2.177 \\ 
			& ESE & 0.402 & 0.187 & 0.368 & 0.468 & 0.284 & 0.228 & 0.225 \\ 
			& Lower CI95 & 25.936 & 1.153 & -0.662 & -2.621 & -1.071 & -0.569 & 1.736 \\ 
			& Upper CI95 & 27.509 & 1.886 & 0.780 & -0.787 & 0.044 & 0.324 & 2.617 \\ 
			\hline
			\multicolumn{9}{l}{\textit{Proposed Methods (Normal Error Assumption)}} \\
			$\tilde{\boldsymbol{\beta}}_{ls}^*$ & Est & 28.083 & 1.949 & 0.897 & -1.107 & -0.349 & -0.118 & 4.012 \\ 
			& ESE & 0.496 & 0.242 & 0.469 & 0.567 & 0.343 & 0.273 & 0.269 \\ 
			& Lower CI95 & 27.111 & 1.474 & -0.023 & -2.219 & -1.020 & -0.653 & 3.485 \\ 
			& Upper CI95 & 29.056 & 2.424 & 1.817 & 0.004 & 0.323 & 0.418 & 4.539 \\ 
			$\tilde{\boldsymbol{\beta}}_{eff}^*$ & Est & 26.793 & 1.385 & 0.558 & -1.141 & -0.372 & -0.136 & 2.931 \\ 
			& ESE & 0.401 & 0.183 & 0.365 & 0.475 & 0.272 & 0.216 & 0.241 \\ 
			& Lower CI95 & 26.007 & 1.027 & -0.158 & -2.072 & -0.905 & -0.559 & 2.460 \\ 
			& Upper CI95 & 27.579 & 1.743 & 1.274 & -0.211 & 0.162 & 0.286 & 3.403 \\ 
			$\widehat{\boldsymbol{\beta}}^*$ & Est & 26.918 & 1.415 & 0.021 & -1.612 & -0.427 & -0.202 & 2.132 \\ 
			& ESE & 0.397 & 0.175 & 0.354 & 0.471 & 0.269 & 0.211 & 0.221 \\ 
			& Lower CI95 & 26.140 & 1.071 & -0.672 & -2.534 & -0.954 & -0.616 & 1.699 \\ 
			& Upper CI95 & 27.696 & 1.758 & 0.715 & -0.689 & 0.101 & 0.211 & 2.566 \\ 
			\hline
			\multicolumn{9}{l}{\textit{Benchmarks}} \\
			$\widehat{\boldsymbol{\beta}}_{bm1}$ & Est & -1.219 & -0.917 & 0.254 & 0.468 & 0.118 & -0.011 & -1.473 \\ 
			& ESE & 0.137 & 0.062 & 0.110 & 0.140 & 0.083 & 0.067 & 0.076 \\ 
			& Lower CI95 & -1.487 & -1.039 & 0.038 & 0.193 & -0.045 & -0.143 & -1.622 \\ 
			& Upper CI95 & -0.951 & -0.796 & 0.470 & 0.743 & 0.280 & 0.121 & -1.323 \\ 
			$\widehat{\boldsymbol{\beta}}_{bm2}$ & Est & 1.076 & -0.687 & 0.356 & 0.386 & 0.112 & -0.064 & -0.898 \\ 
			& ESE & 0.024 & 0.005 & 0.016 & 0.026 & 0.009 & 0.006 & 0.007 \\ 
			& Lower CI95 & 1.029 & -0.697 & 0.325 & 0.336 & 0.096 & -0.075 & -0.913 \\ 
			& Upper CI95 & 1.124 & -0.678 & 0.386 & 0.436 & 0.129 & -0.053 & -0.884 \\ 
			\hline
		\end{tabular}
	}
	\label{tab:bmi}
\end{table}

\label{lastpage}

\end{document}


	
	\date{{\it Received October} 2007. {\it Revised February} 2008.  {\it
			Accepted March} 2008.}
	
	
	
	\pagerange{\pageref{firstpage}--\pageref{lastpage}} 
	\volume{64}
	\pubyear{2008}
	\artmonth{December}
	
	
	\doi{10.1111/j.1541-0420.2005.00454.x}
	
	
	\label{firstpage}
	

	\maketitle
	
	
	%
	
\section{Verification of nonidentifiability in Remark \ref{rem:iden}}\label{sec:iden}
Let $F(t,\x),\bb$ denote the true function and parameter that
describe the data generation process for $Z,\x$.
Let
\bse
s(\x)=I(\ba\trans\x\ne
c)\frac{\Phi^{-1}\{F(c-\bb\trans\x,\x)\}}{c-\ba\trans\x},
\ese
where $\ba\ne\bb$ and $\Phi(\cdot)$ is the standard normal
distribution function.
Let $G(t,\x)=\Phi\{ts(\x)\}$. The $G(t,\x)$ is valid distribution
function with mean zero, and
\bse
G(c-\ba\trans\x,\x)
=\Phi\{(c-\ba\trans\x)s(\x)\}
=F(c-\bb\trans\x,\x).
\ese
Thus, $G(t,\x), \ba$ would describe the same data generation process
hence the model is not identifiable.

\section{Proof of Proposition~\ref{pro:decomp}}\label{app:pro:decomp}

The likelihood function of model (\ref{eq:model}) for the combined population is semiparametric.
Besides the parameter of interest $\bb$, there are three nonparametric nuisance components $f_t(\x)$, $f_e(\x)$ and $f(\epsilon,\x)$, and another nuisance $\pi$.

According to the semiparametric theory \citep{bickel1993efficient, tsiatis2006semiparametric}, influence functions belong to the linear space orthogonal to the nuisance tangent space $\Lambda$, denoted as $\Lambda^\perp$ throughout.
Here, the nuisance tangent space $\Lambda$ is defined as the mean squared closure of nuisance tangent spaces associated with all parametric submodels.

Below, we show the mutually orthogonal decomposition of the space $\calH$ that contains all the $p$-dimensional mean zero measurable functions of the observed data with finite variance.
It is
\bse
\calH = \Lambda_\pi\oplus\Lambda_t \oplus \Lambda_e \oplus \Lambda_f \oplus \Lambda^\perp,
\ese
where
\bse
\Lambda_\pi&=&\left\{\a\left(\frac{r}{\pi}-\frac{1-r}{1-\pi}\right)\right\},\\
\Lambda_t&=&\{r\a(\x): E_t(\a)=\0\},\\
\Lambda_e&=&\{(1-r)\a(\x): E_e(\a)=\0\},\\
\Lambda_f&=&
\{r\a(\epsilon,\x)+(1-r)\frac{E\{Z\a(\epsilon,\x)\mid\x\}}{F(c-\bb\trans\x,\x)}
\frac{z-F(c-\bb\trans\x,\x)}{1-F(c-\bb\trans\x,\x)}:E(\a\mid\x)=\0,E(\epsilon\a\mid\x)=\0\},
\ese
are the nuisance tangent spaces with respect to $\pi$, $f_t(\x)$,
$f_e(\x)$, and $f(\epsilon,\x)$, respectively,
where the notation $\oplus$ represents the direct sum of two spaces that are orthogonal to each other.

The loglikelihood of (\ref{eq:model}) is
\bse
&&r\log f(y-\bb\trans\x,\x)+(1-r)[z\log F(c-\bb\trans\x,\x)+(1-z)\log
\{1-F(c-\bb\trans\x,\x)\}] \\
&&+r\log f_t(\x)+(1-r)\log
f_e(\x)+r\log\pi+(1-r)\log (1-\pi),
\ese
Let $\epsilon=Y-\bb\trans\X$,
\bse
\Lambda_\pi&=&\left\{\a\left(\frac{r}{\pi}-\frac{1-r}{1-\pi}\right)\right\},\\
\Lambda_t&=&\{r\a(\x): E_t(\a)=\0\},\\
\Lambda_e&=&\{(1-r)\a(\x): E_e(\a)=\0\},\\
\Lambda_f&=&\{r\a(\epsilon,\x)+(1-r)z\frac{E\{Z\a(\epsilon,\x)\mid\x\}}{F(c-\bb\trans\x,\x)}
-(1-r)(1-z)\frac{E\{Z\a(\epsilon,\x)\mid\x\}}{1-F(c-\bb\trans\x,\x)}:\\
&&E(\a\mid\x)=\0,E(\epsilon\a\mid\x)=\0\}\\
&=&\{r\a(\epsilon,\x)+(1-r)\frac{E\{Z\a(\epsilon,\x)\mid\x\}}{F(c-\bb\trans\x,\x)}
\frac{z-F(c-\bb\trans\x,\x)}{1-F(c-\bb\trans\x,\x)}:E(\a\mid\x)=\0,E(\epsilon\a\mid\x)=\0\}.
\ese
Obviously, $\Lambda=\Lambda_\pi\oplus\Lambda_t\oplus\Lambda_e\oplus\Lambda_f$.

For $\Lambda^\perp$, we consider
$r\b_1(\epsilon,\x)+(1-r)\{z\b_2(\x)+(1-z)\b_3(\x)\}$ so that
\bse
&&E[R\b_1(\epsilon,\X)+(1-R)\{Z\b_2(\X)+(1-Z)\b_3(\X)\}]=0,\\
&&E[R\b_1(\epsilon,\X)/\pi-(1-R)\{Z\b_2(\X)+(1-Z)\b_3(\X)\}/(1-\pi)]=0,\\
&&E\{R\b_1(\epsilon,\X)\mid\X\}=\c_1,\\
&&E[(1-R)\{Z\b_2(\X)+(1-Z)\b_3(\X)\}\mid\X]=\c_2,
\ese
and
\bse
&&E\left([R\b_1(\epsilon,\X)+(1-R)\{Z\b_2(\X)+(1-Z)\b_3(\X)\}]\trans\right.\\
&&\left.\times[R\a(\epsilon,\x)+(1-R)\frac{E \{Z\a(\epsilon,\x)\mid\X\}}{F( c-\bb\trans\X,\x)}
\frac{z-F( c-\bb\trans\X,\x)}{1-F( c-\bb\trans\X,\x)}]\right)\\
&=&E[R\b_1(\epsilon,\X)\trans\a(\epsilon,\x)+(1-R)\{Z\b_2(\X)+(1-Z)\b_3(\X)\}\trans \\
&& \times \frac{E \{Z\a(\epsilon,\x)\mid\X\}}
{F( c-\bb\trans\X,\x)}\frac{Z-F( c-\bb\trans\X,\x)}{1-F( c-\bb\trans\X,\x)}]\\
&=&E[p(\x)\b_1(\epsilon,\X)\trans\a(\epsilon,\x)
+ (1-p(\x))\{\b_2(\X)-\b_3(\X)\}\trans
E \{Z\a(\epsilon,\x)\mid\X\}
]\\
&=&0.
\ese
Writing $F(c-\bb\trans\x,\x)$ as $F$, $f(c-\bb\trans\x,\x)$ as $f$,
this is
equivalent to
\bse
&& \pi E_t\{\b_1(\epsilon,\X)\}+(1-\pi) E_e[\{F\b_2(\X)+(1-F)\b_3(\X)\}]=\0,\\
&&\c_1/\pi-\c_2/(1-\pi)=0,\\
&&p(\x)E\{\b_1(\epsilon,\X)\mid\X\}=\c_1,\\
&&\{1-p(\x)\}\{F\b_2(\X)+(1-F)\b_3(\X)\}=\c_2,\\
&& p(\x)\b_1(\epsilon,\X)+\{1-p(\x)\}Z\{\b_2(\X)-\b_3(\X)\}
=\c_3(\x)+\c_4(\x)\epsilon.
\ese
This is further equivalent to
\bse
&&\c_1/\pi-\c_2/(1-\pi)=0,\\
&&\b_2(\x)=
\frac{\c_2-\{1-p(\x)\}\b_3(\x)}{\{1-p(\x)\}F}+\b_3(\x),\\
&& \b_1(\epsilon,\X)=
(F-Z)\left[
\frac{\c_2-\{1-p(\x)\}\b_3(\x)}{p(\x) F}\right]
+\frac{\c_4(\x) \epsilon-\c_2}{p(\x)},\\
&&\c_1=-\c_2.
\ese
Hence, $\c_1=\c_2=\0$, and
\bse
&&\b_2(\x)=
\frac{(F-1)\b_3(\x)}{F},\\
&& \b_1(\epsilon,\X)=\frac{\c_4(\x) \epsilon}{p(\x)}-
(F-Z)
\frac{\{1-p(\x)\}\b_3(\x)}{p(\x) F}.
\ese
Thus, we obtain
\bse
\Lambda^\perp
&=&\left(r\left[
\frac{\c(\x) \epsilon}{p(\x)}-
\frac{(F-z)\{1-p(\x)\}\b(\x)}{p(\x) F}
\right]\right.\\
&&\left.+(1-r)\left\{z
\frac{(F-1)\b(\x)}{F}
+(1-z)\b(\x)\right\}:\forall  \c(\x), \b(\x)\right).
\ese
Finally, this space can be even further simplified as
\bse
\Lambda^\perp = \left\{ r(y-\bb\trans\x)\c(\x) + \{r-p(\x)\}\{z-F(c-\bb\trans\x,\x)\}\b(\x): \forall \c(\x),\b(\x) \right\},
\ese
which completes the proof.

\section{Proof of Proposition~\ref{pro:eif}}\label{app:pro:eif}

The score vector is
\bse
\S_\bb&=&-\x\left\{ r\frac{f'(\epsilon,\x)}
{f(\epsilon,\x)}+(1-r)z\frac{f( c-\bb\trans\x,\x)}
{F( c-\bb\trans\x,\x)}-(1-r)(1-z)\frac{f( c-\bb\trans\x,\x) }
{1-F( c-\bb\trans\x,\x)}\right\}\\
&=&-\x\left[ r\frac{f'(\epsilon,\x)}{f(\epsilon,\x)}+(1-r)
\frac{f( c-\bb\trans\x,\x)}{F( c-\bb\trans\x,\x)}
\frac{z-F( c-\bb\trans\x,\x)}
{1-F( c-\bb\trans\x,\x)}\right]\\
&=&-\x\left[ r\frac{f'(\epsilon,\x)}{f(\epsilon,\x)}+(1-r)
\frac{f}{F}
\frac{z-F}
{1-F}\right].
\ese

We can verify that
\be\label{eq:seff}
\S\eff&=&
\frac{[r \epsilon(F-1)F -r (F-z)\{1-p(\x)\}E(Z\epsilon\mid\x)
	+p(\x)(1-r) (F-z) E(Z\epsilon\mid\x)]
	\x}{\{E(Z\epsilon\mid\x)\}^2\{1-p(\x)\}+F(F-1)
	E(\epsilon^2\mid\x)}\n\\
&=&\frac{[r \epsilon(F-1)F
	+\{p(\x)-r\} (F-z) E(Z\epsilon\mid\x)]
	\x}{\{E(Z\epsilon\mid\x)\}^2\{1-p(\x)\}+F(F-1)
	E(\epsilon^2\mid\x)},
\ee
corresponding to $\c=\0$ and
\bse
\c_1(\x)&=&\frac{(F-1)\b(\x)}{E(Z\epsilon\mid\x)},\\
\b(\x)&=&\frac{F
	E(Z\epsilon\mid\x)p(\x)\x}{\{E(Z\epsilon\mid\x)\}^2\{1-p(\x)\}+F(F-1)
	E(\epsilon^2\mid\x)}
\ese
in $\Lambda^\perp$,
and
\bse
\S_\bb-\S\eff
&=&
\frac{-r\x f'(\epsilon,\x)}{f(\epsilon,\x)}
-\frac{(1-r) (z-F) f\x}{F(1-F)}\\
&&-\frac{[r \epsilon(F-1)F
	+\{p(\x)-r\} (F-z) E(Z\epsilon\mid\x)]
	\x}{\{E(Z\epsilon\mid\x)\}^2\{1-p(\x)\}+F(F-1)
	E(\epsilon^2\mid\x)}\\
&=&-r\x\left[\frac{f'(\epsilon,\x)}{f(\epsilon,\x)}
+\frac{ \epsilon(F-1)F -(F-z)\{1-p(\x)\}E(Z\epsilon\mid\x)
}{\{E(Z\epsilon\mid\x)\}^2\{1-p(\x)\}+F(F-1)
	E(\epsilon^2\mid\x)}\right]\\
&&-(1-r)\x\left[f-\frac{p(\x) F(1-F) E(Z\epsilon\mid\x)
}{\{E(Z\epsilon\mid\x)\}^2\{1-p(\x)\}+F(F-1)
	E(\epsilon^2\mid\x)}\right]\frac{z-F}{F(1-F)}\in\Lambda_f.
\ese

\section{Proof of Theorem \ref{th:seff}}\label{app:proof:theorem}
To simplify notation, we
denote $\wt\epsilon_i^*=y_i-\wt\bb\eff^{*\rm T}\x_i$,
$\wt c_i^*=c-\wt\bb\eff^{*\rm T}\x_i$,
$\epsilon_i^*=y_i-\bb^{*\rm T}\x_i$,
$c_i^*=c-\bb^{*\rm T}\x_i$,
$c_i=c-\bb\trans\x_i$,
where
$\bb^*=\lambda\bb+(1-\lambda)\wt\bb\eff^*$ for some $0\le
\lambda\le1$. Let $\bp_\ba'(\x,\ba)=\partial p(\x,\ba)/\partial\ba$.
We first have
\be\label{eq:dp}
&&E\left[\frac{\partial\S\eff^*\{R,RY,Z,\x,\bb,p(\x)\}}{\partial p(\x)}\mid\x\right]\n\\
&=&\frac{\{F^*-E(Z\mid\x)\} E^*(Z\epsilon\mid\x)\x}{\{E^*(Z\epsilon\mid\x)\}^2\{1-p(\x)\}+F^*(F^*-1)
	E^*(\epsilon^2\mid\x)}\n\\
&&+ \left( \frac{[E(R\mid\x)E(\epsilon\mid\x)(F^*-1)F^*
	+\{p(\x)-E(R\mid\x)\} \{F^*-E(Z\mid\x)\} E^*(Z\epsilon\mid\x)]
	\x}{[\{E^*(Z\epsilon\mid\x)\}^2\{1-p(\x)\}+F^*(F^*-1)
	E^*(\epsilon^2\mid\x)]^2} \right. \n \\ && \left. \qquad \times \{E^*(Z\epsilon\mid\x)\}^2 \right)  \n \\
&=&\frac{(F^*-F) E^*(Z\epsilon\mid\x)\x}{\{E^*(Z\epsilon\mid\x)\}^2\{1-p(\x)\}+F^*(F^*-1)
	E^*(\epsilon^2\mid\x)}.
\ee
Next, because $\wt\bb\eff^*$ solves the estimating equation formed by the
summation of the local efficient scores in \eqref{eq:seffs}, we have

\bse
\0&=&N^{-1/2}\sumI\frac{[r_i \wt\epsilon_i^*\{F^*(\wt c_i^*,\x_i)-1\}F^*(\wt c_i^*,\x_i)
	+\{\wh p(\x_i)-r_i\} \{F^*(\wt c_i^*,\x_i)-z_i\}E^*(Z_i\wt\epsilon_i^*\mid\x_i)]
	\x_i}{\{E^*(Z_i\wt\epsilon_i^*\mid\x_i)\}^2\{1-\wh p(\x_i)\}+F^*(\wt c_i^*,\x_i) \{F^*(\wt c_i^*,\x_i)-1\}
	E^*(\wt\epsilon_i^{*2}\mid\x_i)}\\
&=&N^{-1/2}\sumI\frac{[r_i \epsilon_i\{F^*(c_i,\x_i)-1\}F^*(c_i,\x_i)
	+\{\wh p(\x_i)-r_i\} \{F^*(c_i,\x_i)-z_i\}E^*(Z_i\epsilon_i\mid\x_i)]
	\x_i}{\{E^*(Z_i\epsilon_i\mid\x_i)\}^2\{1-\wh p(\x_i)\}+F^*(c_i,\x_i) \{F^*(c_i,\x_i)-1\}
	E^*(\epsilon_i^2\mid\x_i)}\\
&&+\left(N^{-1}\sumI \frac{\partial}{\partial\bb\trans}\frac{[r_i \epsilon_i^*\{F^*(c_i^*,\x_i)-1\}F^*(c_i^*,\x_i)
	+\{\wh p(\x_i)-r_i\} \{F^*(c_i^*,\x_i)-z_i\}E^*(Z_i\epsilon_i^*\mid\x_i)]
	\x_i}{\{E^*(Z_i\epsilon_i^*\mid\x_i)\}^2\{1-\wh p(\x_i)\}+F^*(c_i^*,\x_i) \{F^*(c_i^*,\x_i)-1\} E^*(\epsilon_i^{*2}\mid\x_i)}\right)\\
&&\times N^{1/2}(\wt\bb\eff^*-\bb)\\
&=&N^{-1/2}\sumI\S\eff^*\{r_i,r_iy_i,z_i,\x_i,\bb,p(\x_i,\wh\ba)\} \\
&& +\left[E\left\{\frac{\partial\S\eff^*(R,RY,Z,\X,\bb,p)}{\partial\bb\trans}\right\}
+o_p(\pi)\right] N^{1/2}(\wt\bb\eff^*-\bb)\\
&=&N^{-1/2}\sumI\S\eff^*\{r_i,r_iy_i,z_i,\x_i,\bb,p(\x_i)\} \\
&& +N^{-1/2}\sumI\frac{\partial\S\eff^*\{r_i,r_iy_i,z_i,\x_i,\bb,p(\x_i,\ba)\}}{\partial\ba\trans}
(\wh\ba-\ba)+o_p(1)\\
&&+\left[E\left\{\frac{\partial\S\eff^*(R,RY,Z,\X,\bb)}{\partial\bb\trans}\right\}  +o_p(\pi)\right] N^{1/2}(\wt\bb\eff^*-\bb)\\
&=&N^{-1/2}\sumI\S\eff^*\{r_i,r_iy_i,z_i,\x_i,\bb,p(\x_i)\} \\
&& +\left(E\left[\frac{\partial\S\eff^*\{R,RY,Z,\X_i,\bb,p(\X)\}}{\partial
	p(\X)}\bp_\ba'(\X,\ba)\trans\right]+o_p(\pi)\right)\\
&&\times \left\{N^{-1/2}
\sumI\bphi(r_i,\x_i, p(\cdot),\ba)+o_p(\pi^{1/2})\right\}
+\left\{\pi\A +o_p(\pi)\right\} N^{1/2}(\wt\bb\eff^*-\bb),
\ese
therefore
$n^{1/2}(\wt\bb\eff^*-\bb)\sim N(\0, \A^{-1}\B{\A^{-1}}\trans)$ in distribution when
$n\to\infty$, where
\bse
\A&=&E(\pi^{-1}\partial\S\eff^*/\partial\bb\trans)\\
&=&E\left(
\frac{R\X}{\pi}\frac{\partial}{\partial\bb\trans}
\left[\frac{ \epsilon(F^*-1)F^*- (F^*-Z) E^*(Z\epsilon\mid\X)]
}{\{E^*(Z\epsilon\mid\X)\}^2\{1-p(\X)\}+F^*(F^*-1)
	E^*(\epsilon^2\mid\X)}\right]\right)\\
&&+E\left(\frac{p(\X)\X}{\pi}\frac{\partial}{\partial\bb\trans}\left[\frac{(F^*-Z) E^*(Z\epsilon\mid\X)
}{\{E^*(Z\epsilon\mid\X)\}^2\{1-p(\X)\}+F^*(F^*-1)
	E^*(\epsilon^2\mid\X)}\right]\right)\\
&=&E\left\{
\frac{p(\X)\X}{\pi}E\left(\frac{\partial}{\partial\bb\trans}
\left[\frac{ \epsilon(F^*-1)F^*- (F^*-Z) E^*(Z\epsilon\mid\X)]
}{\{E^*(Z\epsilon\mid\X)\}^2+F^*(F^*-1)
	E^*(\epsilon^2\mid\X)}\right]\mid\X\right)\right\}\\
&&+E\left\{\frac{p(\X)\X}{\pi}E\left(\frac{\partial}{\partial\bb\trans}\left[\frac{(F^*-Z) E^*(Z\epsilon\mid\X)
}{\{E^*(Z\epsilon\mid\X)\}^2+F^*(F^*-1)
	E^*(\epsilon^2\mid\X)}\right]\mid\X\right)\right\}\\
&=&E\left\{
\frac{p(\X)\X}{\pi}E\left(\frac{\partial}{\partial\bb\trans}
\left[\frac{ \epsilon(F^*-1)F^*
}{\{E^*(Z\epsilon\mid\X)\}^2+F^*(F^*-1)
	E^*(\epsilon^2\mid\X)}\right]\mid\X\right)\right\},
\ese
and
\bse
\B&=&E\left\{\frac{1}{\pi}\left(\S\eff^*\{R,RY,Z,\X,\bb,p(\X)\} \right.\right.\\
&&\left.\left. +E\left[\frac{\partial\S\eff^*\{R,RY,Z,\X,\bb,p(\X)\}}{\partial
	p(\X)}\bp_\ba'(\X,\ba)\trans\right]
\bphi(R,\X, p(\cdot),\ba)\right)^{\otimes2}\right\}\\
&=&E\left\{\frac{1}{\pi}\left(\S\eff^*\{R,RY,Z,\X,\bb,p(\X)\}\right.\right.\\
&&\left.\left.+E\left[ \frac{(F^*-F) E^*(Z\epsilon\mid\x) \x\bp_\ba'(\X,\ba)\trans}{\{E^*(Z\epsilon\mid\x)\}^2\{1-p(\x)\}+F^*(F^*-1)
	E^*(\epsilon^2\mid\x)}\right]
\bphi(R,\X, p(\cdot),\ba)\right)^{\otimes2}\right\}.
\ese


\section{Proof of the Guaranteed Efficient Gain of $\wh\bb^*$} \label{app:proof:eff_gain}
In Section~\ref{eff_gain}, we proposed estimator $\wh\bb^*$ which is guaranteed to be more efficient than $\wt\bb\ls^*$. It is because that
\bse
\var(\wt\bb\ls^*)-\var(\wh\bb^*)
=\{\cov(\wt\bb\ls^*,
\wt\bb\eff^*-\wt\bb\ls^*)\}\{\var(\wt\bb\eff^*-\wt\bb\ls^*)\}^{-1}
\{\cov(\wt\bb\eff^*-\wt\bb\ls^*, \wt\bb\ls^*)\},
\ese
is positive definite. 
Of course, the estimator is also always more efficient than
$\wt\bb\eff^*$ in that we also have
\bse
\var(\wt\bb\eff^*)-\var(\wh\bb^*)
=\{\cov(\wt\bb\eff^*,
\wt\bb\eff^*-\wt\bb\ls^*)\}\{\var(\wt\bb\eff^*-\wt\bb\ls^*)\}^{-1}
\{\cov(\wt\bb\eff^*-\wt\bb\ls^*, \wt\bb\eff^*)\},
\ese
which is also positive definite.

Additionally, if the working model $f^*(\epsilon,\x)$ is fortunately
chosen as the correct model, then the matrix $\W$ becomes the identity
matrix, and the estimator $\wh\bb^*$ (now written as $\wh\bb$) becomes
$\wt\bb\eff$, the semiparametrically efficient estimator based on the
efficient score $\S\eff$. Hence, the estimator $\wh\bb^*$ also guarantees the efficient use of
the external data. 

In practice, $\W$ is unknown and has to be
estimated. As long as a consistent estimator $\wh\W$ is used, the
resulting $\wt\bb\ls^* + \wh\W (\wt\bb\eff^*-\wt\bb\ls^*)$ is still
optimal because we can verify that 
$\var\{\wt\bb\ls^* + \wh\W (\wt\bb\eff^*-\wt\bb\ls^*)\}
=\var\{\wt\bb\ls^* + \W (\wt\bb\eff^*-\wt\bb\ls^*) +o_p(1)
(\wt\bb\eff^*-\wt\bb\ls^*)\}
=\var(\wh\bb^*)+o(1)$ and $E\{\wt\bb\ls^* + \wh\W (\wt\bb\eff^*-\wt\bb\ls^*)\}
=E\{\wt\bb\ls^* + \W (\wt\bb\eff^*-\wt\bb\ls^*) +o_p(1)
(\wt\bb\eff^*-\wt\bb\ls^*)\}
=E(\wh\bb^*)+o(1)$.

\section{Estimation Accuracy of the Weighting Matrix $\W$ in $\wh\bb^*$}\label{app:simulation:W}

To verify the estimation accuracy of the weighting matrix $\W$ used in the proposed estimator $\wh\bb^*$, we conduct additional simulation studies under the settings of Simulation 1 Part 1 Scenario III, considering both small and large sample sizes, to asssess how the weighted bootstrap estimator $\W_{bt}$ approximates its theoretical value $\W_t$.


As summarized in Table~\ref{tab:ssd_sse_ci}, the results for all three
regression coefficients show that the sample standard deviation (SSD),
average estimated standard error (ESE), and coverage probabilities of
the bootstrap-based estimator are nearly indistinguishable from those
based on the true weighting matrix, even in the small-sample case
($n=500$). This finding suggests that reliance on the bootstrap estimator does not compromise estimation efficiency and continues to
deliver reliable inference. 

Complementary evidence comes from the direct comparisons of $\W_{bt}$
and $\W_{t}$. Table~\ref{tab:elementwise} shows that the mean and
median elementwise differences between the two matrices, which can
reach magnitudes of 0.10 when $n=500$, shrink markedly toward zero as
the sample size increases to $n=2400$. Consistent with this trend, the
Frobenius norm results in Table~\ref{tab:frobenius} confirm an
improvement in overall matrix accuracy, with the mean discrepancy
decreasing from 0.58 at $n=500$ to 0.37 at $n=2400$. Scaling this
discrepancy relative to the size of the true weighting matrix
(Table~\ref{tab:relative_frobenius}) further highlights the
improvement, as the average relative error declines from about 26\% in
small samples to only 17\% in larger samples. 

Overall, these results demonstrate that bootstrap estimation of
the weighting matrix provides a close 
approximation to the true weighting matrix, and that the accuracy of this
approximation improves substantially as the sample size grows.

\begin{table}[htbp]
	\centering
	\caption{
		Comparsion of the corresponding estimators $\wh\bb^*$, based on either $\W_t$ or $\W_{bt}$, in Simulation 1 Part 1 Scenario III.
		}
	\begin{tabular}{lcccccc}
		\hline
		& & \multicolumn{2}{c}{$n=500$} & \multicolumn{2}{c}{$n=2400$} \\
		 &  & $\W_{t}$ & $\W_{bt}$ & $\W_{t}$ & $\W_{bt}$ \\
		\hline
		$\beta_{0}$ & SSD & 0.3437 & 0.3428 & 0.1966 & 0.1950 \\
		& ESE & 0.3198 & 0.3279 & 0.1896 & 0.1921 \\
		& CR95  & 0.9127 & 0.9310 & 0.9279 & 0.9350 \\
		\hline
		$\beta_{1}$ & SSD & 0.4959 & 0.4953 & 0.2968 & 0.2942 \\
		& ESE & 0.4707 & 0.4740 & 0.2820 & 0.2825 \\
		& CR95 & 0.9278 & 0.9330 & 0.9279 & 0.9340 \\
		\hline
		$\beta_{2}$ & SSD & 0.2434 & 0.2416 & 0.1531 & 0.1502 \\
		& ESE & 0.2374 & 0.2381 & 0.1424 & 0.1422 \\
		& CR95  & 0.9378 & 0.9450 & 0.9289 & 0.9370 \\
		\hline
	\end{tabular}
	\label{tab:ssd_sse_ci}
\end{table}

\begin{table}[htbp]
	\centering
	\caption{
		Comparison between $\W_{t}$ and $\W_{bt}$, in elementwise mean and median difference, in Simulation 1 Part 1 Scenario III.
		}
	\begin{tabular}{lrrrr}
		\hline
		Entry & Mean ($n=500$) & Mean ($n=2400$) & Median ($n=500$) & Median ($n=2400$) \\
		\hline
		(1,1) &  0.1075 &  0.0147 &  0.1023 &  0.0109 \\
		(1,2) &  0.0061 &  0.0059 &  0.0064 &  0.0068 \\
		(1,3) & -0.0074 & -0.0050 &  0.0149 &  0.0052 \\
		(2,1) & -0.0647 & -0.0199 & -0.0843 & -0.0228 \\
		(2,2) &  0.0356 & -0.0049 &  0.0348 &  0.0014 \\
		(2,3) &  0.0160 &  0.0104 & -0.0133 & -0.0032 \\
		(3,1) &  0.0875 &  0.0409 &  0.0884 &  0.0403 \\
		(3,2) &  0.0502 &  0.0247 &  0.0484 &  0.0227 \\
		(3,3) &  0.0919 & -0.0014 &  0.0739 &  0.0145 \\
		\hline
		\label{tab:elementwise}
	\end{tabular}
\end{table}

\begin{table}[htbp]
	\centering
	\caption{
		Comparison between $\W_{t}$ and $\W_{bt}$, in Frobenius norm, in Simulation 1 Part 1 Scenario III.
		}
	\begin{tabular}{lcccccc}
		\hline
		Sample size & Min & $Q_1$ & Median & Mean & $Q_3$ & Max \\
		\hline
		$n=500$  & 0.093 & 0.403 & 0.543 & 0.583 & 0.717 & 2.283 \\
		$n=2400$ & 0.097 & 0.258 & 0.344 & 0.371 & 0.460 & 1.140 \\
		\hline
		\label{tab:frobenius}
	\end{tabular}
\end{table}

\begin{table}[htbp]
	\centering
	\caption{
		Comparison between $\W_{t}$ and $\W_{bt}$, in relative Frobenius norm, defined as $\|\W_{t}-\W_{bt}\|_{F}/\|\W_{t}\|_{F}$, in Simulation 1 Part 1 Scenario III.
		}
	\begin{tabular}{lcccccc}
		\hline
		Sample size & Min & $Q_1$ & Median & Mean & $Q_3$ & Max \\
		\hline
		$n=500$  & 0.049 & 0.180 & 0.237 & 0.260 & 0.316 & 1.016 \\
		$n=2400$ & 0.043 & 0.118 & 0.157 & 0.170 & 0.209 & 0.562 \\
		\hline
		\label{tab:relative_frobenius}
	\end{tabular}
\end{table}

	\backmatter

	\bibliographystyle{biom} 
	\bibliography{refMIXEDtransfer.bib}

	\label{lastpage}